\documentclass{aa}  

\usepackage{graphicx}
\usepackage{txfonts}
\usepackage{hyperref} 
\usepackage{float} 

\begin{document}

\title{Imaging a ring-like structure and the extended jet of M87 at 86\,GHz}
\titlerunning{Imaging the ring and extended jet of M87}
\authorrunning{Kim et. al.}


\author{Jong-Seo Kim \inst{1}\fnmsep\thanks{jongkim@mpifr-bonn.mpg.de}
\and Hendrik Müller \inst{1,2}
\and Aleksei S. Nikonov \inst{1}
\and Ru-Sen Lu \inst{3,4,1}
\and Jakob Knollmüller \inst{5}
\and Torsten A. En{\ss}lin \inst{6,7}
\and Maciek Wielgus \inst{8,1}
\and Andrei P. Lobanov \inst{1,9}
}

\institute{    
            Max-Planck-Institut f\"ur Radioastronomie, Auf dem H\"ugel 69, D-53121 Bonn, Germany
        \and Jansky Fellow of National Radio Astronomy Observatory, 1011 Lopezville Rd, Socorro, NM 87801, USA
        \and
        Key Laboratory of Radio Astronomy and Technology, Chinese Academy of Sciences, A20 Datun Road, Chaoyang District, Beijing, 100101, People’s Republic of China
        \and  Shanghai Astronomical Observatory, Chinese Academy of Sciences, Shanghai, People’s Republic of China
        \and
        Radboud University, Heyendaalseweg 135, 6525 AJ, Nijmegen, The Netherlands
        \and
        Max-Planck-Institut f\"ur Astrophysik, Karl-Schwarzschild-Str. 1, 85748 Garching, Germany
        \and
        Ludwig-Maximilians-Universit\"at, Geschwister-Scholl-Platz 1, 80539 Munich, Germany
            \and
         Instituto de Astrofísica de Andalucía-CSIC, Glorieta de la Astronomía s/n, E-18008 Granada, Spain 
        \and
        Institut für Experimentalphysik, Universit\"at Hamburg, Luruper Chaussee 149, 22761, Hamburg, Germany  
         }
   \date{Received August 29, 2024; accepted February 24, 2025}

 
  \abstract
   {The galaxy M87 is one of the prime targets for high resolution radio imaging to investigate the ring-like "shadow" of its supermassive black hole, the innermost regions of accretion flow, and the formation of the relativistic jet. However, it remains challenging to observe them jointly due to the sparsity of the UV coverage and limited array sensitivity. In 2018, global mm-VLBI array (GMVA)+ALMA observations at 86\,GHz  enabled the simultaneous reconstruction of a ring structure and the extended jet emission. In order to analyze the ring and jet of M87, conventional \texttt{CLEAN} algorithms were mainly employed alongside the regularized maximum likelihood method \texttt{SMILI} in previous work.}
   {To test the robustness of the reconstructed structures of M87 GMVA+ALMA observations at 86\,GHz, we estimate the ring diameter, width, and the extended jet emission with the possible central spine by two different novel imaging algorithms: \texttt{resolve} and \texttt{DoG-HiT}.}
   {We performed Bayesian self-calibration and imaging with uncertainty estimation using \texttt{resolve}. In addition, we reconstructed the image with \texttt{DoG-HiT}, using only interferometric closure quantities.}
   {Overall, reconstructions are consistent with the \texttt{CLEAN} and \texttt{SMILI} images. The ring structure of M87 is resolved at a higher resolution and the posterior distribution of M87 ring features is explored. The \texttt{resolve} images show that the ring diameter is $60.9 \pm 2.2 \, \mu \mathrm{as}$ and its width is $16.0 \pm 0.9 \, \mu\mathrm{as}$. The ring diameter and the ring width measured from the \texttt{DoG-HiT} image are $61.0 \, \mu \rm{as}$ and $20.6 \, \mu \rm{as}$, respectively. The ring diameter is therefore in agreement with the estimation ($64^{+4}_{-8} \, \mu\mathrm{as}$) by \texttt{SMILI} image reconstructions and visibility domain model fitting. Two bright spots in the ring are reconstructed by four independent imaging methods. Therefore, the substructure in the ring most likely results from the data. A consistent limb-brightened jet structure is reconstructed by \texttt{resolve} and \texttt{DoG-HiT}, albeit with a less pronounced central spine.
   }
   {Modern data-driven imaging methods confirm the ring and jet structure in M87, and complement traditional VLBI methods with novel perspectives on evaluating the significance of the recovered features. They confirm the result of the previous report.}


   \maketitle

\section{Introduction}
\label{Chap1}

Continued improvements in instruments and algorithms for very long baseline interferometry (VLBI) enable us to image and analyze radio emission from near to supermassive black holes (SMBH) and inside the formation zone of relativistic jet in active galactic nuclei (AGNs). The recent Event Horizon Telescope (EHT) observations captured the black hole "shadow" in M87* and SgrA* \citep{EHT2019_M87_paper1, eht2022a}, and the Global mm-VLBI Array (GMVA) and Very Long Baseline Array (VLBA) observations  provided high fidelity images of the extended jet emission in radio-loud AGNs \citep{Walker2018, JYKim2018, Okino_2022}. The nearby galaxy M87, with a SMBH mass of $6.5 \times 10^9 \textrm{M}_{\odot}$ \citep{eht2019_paper6_mass} and a redshift of $z = 0.004283$ \citep{2011MNRAS.413..813C}, 
provides a unique opportunity to study the black hole accretion disk and jet launching mechanism simultaneously due to the large angular scales of 0.08 pc or 260 $\textrm{R}_g$ (where $\textrm{R}_g = GM/c^2$) per milliarcsecond. So far, the black hole shadow and relativistic jet of M87 at radio frequencies have been observed independently due to instrumental limitations and the characteristics of the source in different frequency regimes.

Observations of M87 with the GMVA, the phased Atacama Large Millimeter/Submillimeter Array (ALMA), and the Greenland Telescope (GLT) in 2018 enabled the central ring-like structure and the extended jet to be imaged simultaneously. \cite{Lu2023} employed a \texttt{CLEAN} \citep{Hogbom1974, 1980A&A....89..377C} self-calibration and imaging method in order to reconstruct the ring and jet structures in M87 with a large field of view. Additionally, the ring in M87 was reconstructed with a smaller field of view using a regularized maximum likelihood (RML)-based imaging software \texttt{SMILI} \citep{Akiyama2017}. GMVA observations at 86 GHz play a significant role in resolving the core and extended jet emission of M87. Prior to the EHT M87 observation in 2017 \citep{EHT2019_M87_paper1}, the GMVA M87 observation in 2014 and 2015 \citep{JYKim2018} provided the core and edge-brightened jet of M87 image down to 13 $\textrm{R}_g$. The GMVA observations with ALMA in 2018 facilitated the detection of the ring with extended jet emission since ALMA provides the longest north-south baselines with improved sensitivity. For the first time, these finding connected the central ring-like feature, which presumably corresponds to the compact accretion flow, with the innermost jet. This work was accompanied by recent groundbreaking observations of the jet in M87 on a variety of scales, ranging from the (full polarimetric results on) horizon scales \citep{EHT2019_M87_paper1, eht2021a, eht2023circ, eht2024a}, to the strongly edge-brightened innermost jet \citep{Walker2018, JYKim2018, JYKim2023}, as well as to the triple-peaked, helical structure and dynamics in the large scale jet \citep{2016ApJ...833...56A, 2017Galax...5....2H, Nikonov2023, Cui2023}. It is imperative to connect the compact scale structures of M87* to the large scale structures to understand AGN jet formation. The observations presented by \citet{Lu2023} play a significant role in connecting horizon scales to jet scales. 

However, the data reduction process of GMVA+ALMA is challenging due to the sparsity of the UV coverage, tropospheric phase corruption at millimeter wavelengths, a low signal-to-noise ratio (S/N), and inhomogeneous antenna statistics due to the different sensitivity of antennas. This difficulty jeopardizes the interpretation of some of the features in the reconstruction, namely the ring-like structure (two brighter spots) and the jet (inner ridge line). Imaging with the conventional \texttt{CLEAN} method is not able to reconstruct a robust ring-like structure due to suboptimal resolution. Furthermore, the simple assumption in \texttt{CLEAN} that the sky brightness distribution is a collection of point sources is not an optimal prior to describe the extended continuous jet emission. An independent assessment of the robustness of these features is needed to facilitate a much-awaited scientific interpretation.

The latest advancement in forward modeling imaging algorithms enables us to generate more robust results from sparse millimeter-VLBI data sets. As an example, Bayesian imaging is a probabilistic approach that reconstructs the posterior distribution using Bayes' theorem. Hence, Bayesian imaging is able to use posterior samples to estimate the uncertainty of parameters, such as image features and instrumental gains. However, image reconstruction is computationally demanding compared to other imaging methods since a collection of possible images is reconstructed instead of a single image. RML methods reconstruct an image by minimizing an objective function, which consists of a data fidelity term and regularizers that favor certain image structures such as smoothness or sparsity. Contrary to \texttt{CLEAN}, forward modeling methods fit the model to the data directly in the visibility domain. We can fit closure quantities directly or calibration can be incorporated in the image reconstruction. Furthermore, we are able to encode knowledge about the source and measurement setup explicitly in the prior distribution and regularizers. As a result, both of the imaging approaches outperform the traditional inverse modeling \texttt{CLEAN} algorithm and we can reconstruct reproducible images with improved resolution in a less supervised fashion. A detailed comparison between \texttt{CLEAN} and forward modeling approaches can be found in \cite{Chael2016, Arras_2021_CygA, Mueller2024_STIX_VLBI}. 

In this work, we reconstruct images using two imaging algorithms: we perform Bayesian self-calibration and imaging jointly using the Bayesian imaging software \texttt{resolve} \citep{Junklewitz2016, Arras2022, Roth2023, JSKim2024} and we reconstruct an image with closure amplitudes and closure phases only \citep{Chael2018} using the RML-based \texttt{DoG-HiT} software \citep{Mueller2022, Mueller2023a, Mueller2023b}. These two independent imaging methods are utilized in order to estimate the robustness of the M87 ring and extended jet emission from the GMVA+ALMA observation in 2018. They quantify the robustness of the recovered features from two alternative, supplementary perspectives, by self-calibration and imaging from a probabilistic point of view and by imaging  with closure quantities only without a potential self-calibration bias.

This article is structured as follows. In Sect. 2 we explain the \texttt{resolve} and \texttt{DoG-HiT} image reconstruction methods. In Sect. 3 we show two image reconstruction results and then analyze the robustness of the M87 ring structure and jet emission. In Sect. 4 we summarize our results.

\begin{figure}[h]
    \centering
    \includegraphics[width=9cm]{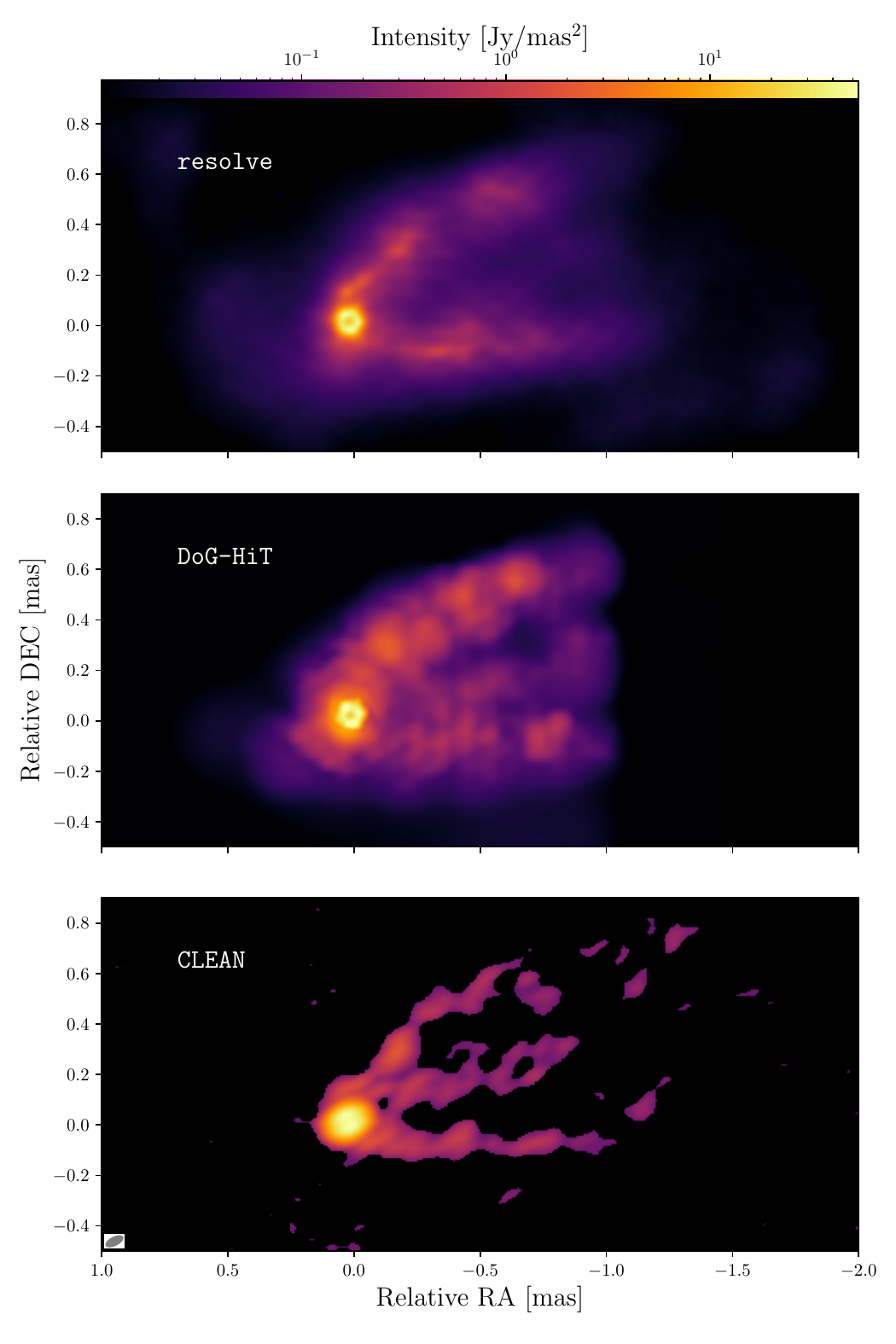}
    \caption{Image reconstructions of GMVA+ALMA M87 at 86\,GHz. Each image presents results obtained by a different algorithm. The top posterior mean image was reconstructed using the \texttt{resolve} Bayesian self-calibration and imaging method. The middle image was obtained using \texttt{DoG-HiT} closure amplitudes and closure phases only imaging. The bottom image was obtained using the \texttt{CLEAN} self-calibration and imaging in \citet{Lu2023}. The \texttt{CLEAN} image is convolved with an elliptical beam, which is represented as an ellipse with sizes  $79 \times 37$\,$\mu$as, P.A.\,$= -63^{\circ}$ in the bottom-left corner.} 
    \label{fig:summary}
\end{figure}

\section{Method}
\label{Chap2}

\subsection{Bayesian self-calibration and imaging by \texttt{resolve}}

\texttt{resolve}\footnote{\url{https://gitlab.mpcdf.mpg.de/ift/resolve}} is an open-source Bayesian imaging software for radio interferometric data \citep{Junklewitz2016, Arras2022, Roth2023, JSKim2024}. In \texttt{resolve}, samples of potential images and antenna-based gain solutions that are consistent with the data are reconstructed by Bayes' theorem in a variational inference sense \citep{Blei2016_VI_review, Knollmueller2019, Frank2021}. In this paper, we used the metric Gaussian variational inference method \citep[][MGVI]{Knollmueller2019} to estimate the posterior distribution of the sky brightness distribution and antenna-based gains. The MGVI method enables us to perform high-dimensional Bayesian inference with affordable computational resources. The probabilistic approach could be advantageous for the M87 GMVA+ALMA observations due to the sparse UV coverage, and large and heterogeneous data uncertainties. For details of  the M87 GMVA+ALMA data, we refer to Sects. 1 and 2 of the supplementary information in \cite{Lu2023}. 

We reconstructed the \texttt{resolve} image in Fig. \ref{fig:summary} with a spatial domain of $2048$ $\times$ $1024$ pixels and a field of view of $4$ mas $\times$ $2$ mas from a-priori calibrated (without self-calibration) data. The Bayesian self-calibration is performed simultaneously with the imaging. The number of posterior samples was 100 and the reduced $\chi^{2}$ value of the final result was $1.1$. The wall-clock time for the \texttt{resolve} reconstruction was $5.5$ hours on a single node of the MPIfR cluster with 25 MPI (Message Passing Interface) tasks.

Since the GMVA+ALMA array is highly inhomogeneous, it is a reasonable assumption that each antenna gain has a different temporal correlation structure. In \texttt{resolve}, we utilized a Gaussian process prior with a nonparametric correlation kernel in the NIFTy software\footnote{\url{https://gitlab.mpcdf.mpg.de/ift/nifty}} for the sky brightness distribution (image) and gain prior model. The spatial correlation between image pixels and temporal correlation between gains can therefore be inferred from the data without manual steering of the gain solution interval constraints (see Fig. \ref{fig:resolve_amp_gain_plot}). The amplitude gain prior is assumed to be correlated and different correlation kernels are inferred per antenna. Gain amplitudes for right-hand circular polarization (RCP) and left-hand circular polarization (LCP) are assumed to have the same correlation structure to stabilize the self-calibration and image reconstruction. The phase gain is assumed to be uncorrelated since the phase coherence time in GMVA observations is comparable with the data averaging time (ten seconds) and it can be even shorter under poor weather conditions.

Furthermore, the posterior distribution of model parameters, such as each pixel in the image and antenna-based gain solutions, can be explored in \texttt{resolve}. In other words, the reliability of the reconstructed parameters and the image features, such as the ring structure and extended jet emission, can be quantified by estimated uncertainties from posterior samples. As an example, if one antenna is problematic, then it would result in a high uncertainty in its gain solution. As a result, in Bayesian imaging, the high uncertainty of these data points can be self-consistently taken into account in the image reconstruction. More details about the Bayesian self-calibration and imaging method for VLBI data and validation with synthetic data can be found in \citet{JSKim2024}.


\subsection{Imaging with closure quantities by \texttt{DoG-HiT}}
\texttt{DoG-HiT} is a regularized maximum likelihood (RML) imaging algorithm that reconstructs the image using  multiscalar wavelet basis functions \citep{Mueller2022}. The basis functions are fitted to the UV coverage, offering a neat separation between covered and noncovered Fourier coefficients, that is, gaps in the UV coverage \citep[for more details on the wavelets, we refer the reader to][]{Mueller2023a}. The image is recovered by a sparsity promoting forward-backward splitting framework that effectively calculates the multiresolution support. In other words, the image is represented by the set of all statistically significant wavelet scales. It has been demonstrated that the multiresolution support is beneficial prior information that enables reconstruction even for sparse and weakly constrained settings \citep{Mueller2022, Mueller2023b}.

In this work, we aim to validate the results presented in \citet{Lu2023} using closure-only imaging pioneered by \citet{Chael2018}. Closure-only imaging is a self-calibration independent technique. As a result, we are able to estimate the robustness of the recovered features against the gain calibration. We represent the \texttt{DoG-HiT} image (see Fig. \ref{fig:summary}) in a square field of view of $4096\,\mu\mathrm{as}$ by $512 \times 512$ pixels. The reconstruction was run on a CPU (11th generation Intel core i7) with 32 GB RAM for roughly thirty minutes. The reduced $\chi^{2}$ to the closure phases was $1.36$ and to the closure amplitudes was $1.1$. For the reconstruction with \texttt{DoG-HiT}, we first select a set of wavelet basis functions, called a dictionary, fitted to the UV coverage of the observation. Then we run \texttt{DoG-HiT} with all large scale wavelets that were significant to fit the extended, diffuse jet emission. Then we use this image as an initial guess and add all small-scale wavelets that are relevant to represent the central ring, and minimize the $\chi^2$-metric to the closure phases and closure amplitudes.

In this framework, we directly fit to the self-calibration independent closure amplitudes and closure phases. For a given number of antennas, more closure quantities can be constructed than visibilities. However, not all the measurements are independent since they can be represented as linear combinations of other closure triangles and quadrilaterals \citep{Twiss1960, Blackburn2020, Thyagarajan2022}. Since the number of statistically independent closure phases and amplitudes is therefore smaller than the number of independent visibilities, which effectively leads to a number of degeneracies such as  lost information of the total flux density and the absolute source position, we have to account for the larger freedom in the models. There are two opposite strategies to achieve this. We could either try to explore the multimodality and degeneracies inherent to closure quantities, for example using recently proposed multiobjective optimization schemes \citep{Mueller2023c,  Mus2024b, Mus2024a}, or we could utilize a more constraining piece of prior information that resolves the degeneracies. For the purpose of the latter approach, multiresolution support proved successful and was implemented in \texttt{DoG-HiT}.

\section{Results}
\label{Chap3}


The M87 GMVA+ALMA images at 86\,GHz by \texttt{resolve}, \texttt{DoG-HiT}, and \texttt{CLEAN} \citep{Lu2023} are shown in Fig. \ref{fig:summary}. \texttt{resolve} and \texttt{DoG-HiT} image fits and results are archived in zenodo \footnote{\url{https://zenodo.org/uploads/13348953}}.

\begin{figure*}[t]
    \centering
    \includegraphics[width=18cm, trim={0cm 0cm 0cm 0cm},clip]{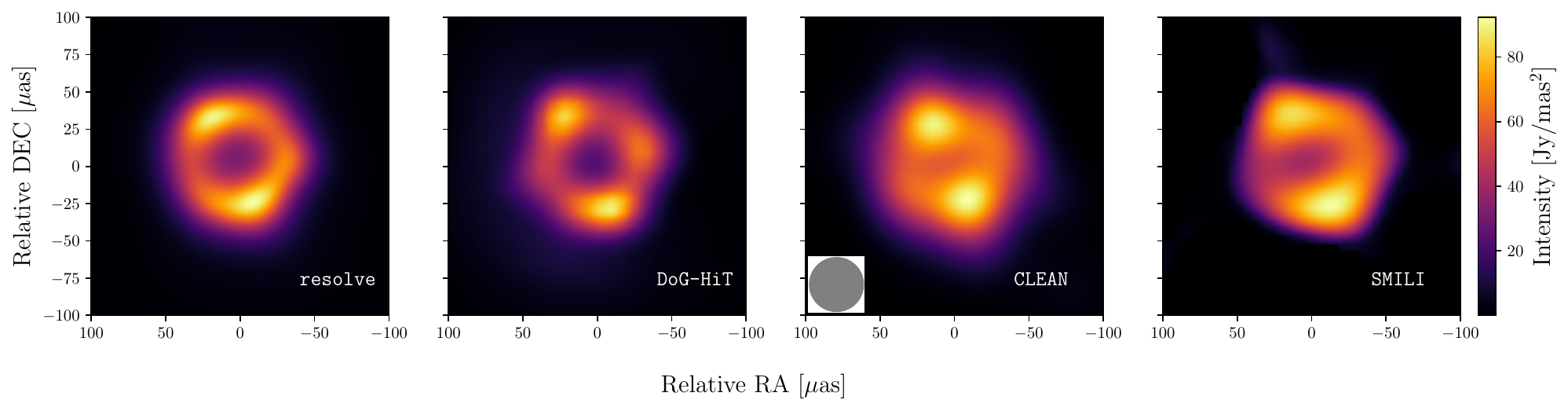}
    \caption{Ring of the M87 image reconstructions at 86\,GHz from  Fig. \ref{fig:summary}. The color shows intensity in Jy~mas$^{-2}$ according to the linear color bar located at the right of the figure. Each image presents results obtained by a different algorithm, whose names are indicated in the lower right corners. The left image is the \texttt{resolve} posterior mean image using the Bayesian self-calibration and imaging method. The left-center image is the \texttt{DoG-HiT} reconstruction using closures only imaging. The right-center image represents the \texttt{CLEAN} reconstruction with the over-resolved 37\,$\mu$as circular beam in \citet{Lu2023}. The right image shows the \texttt{SMILI} reconstruction in \citet{Lu2023}. All images were processed by a Gaussian interpolation.}
    \label{fig:ring_compare}
\end{figure*}

\subsection{Estimation of ring-like features in M87} 

A zoom-in on the central compact emission region is presented in Fig. \ref{fig:ring_compare}. In the visibility domain, the presence of the visibility null amplitude at around 2.3\,G$ \lambda$ and the phase jump around the null amplitude \citep[see Fig. S2 in][]{Lu2023} is analogous to EHT M87 observations in 2017 and 2018 \citep{EHT2019_M87_paper1, eht2024a}, which is strong evidence of the M87 ring structure. Figure \ref{fig:uv_plot} (middle panel) shows the visibility domain representation of \texttt{resolve} and \texttt{DoG-HiT} images, namely the posterior mean model visibility amplitudes by \texttt{resolve} and the amplitudes of the Fourier-transformed \texttt{DoG-HiT} image. The visibility null amplitudes in \texttt{resolve} and \texttt{DoG-HiT} are located at around 2.3\,G$ \lambda$, which is consistent with the results in \cite{Lu2023}. The visibility null amplitudes are shorter than the M87 EHT observation in 2017 and 2018 (at around 3.4\,G$ \lambda$).

The disagreement of visibility null amplitudes between 3\,mm and 1\,mm implies that the diameter of the observed ring at 3\,mm is larger than that of the ring at 1\,mm, since the baseline location of the visibility null amplitude scales inversely with the ring diameter \citep{Lu2023}. For comparison, in Sagittarius A*, the intrinsic (descattered) size of the compact horizon-scale source changes by about a factor of 2 between 1\,mm and 3\,mm emission \citep{Issaoun2019,eht2022a}. 
The M87 ring diameter at 86\,GHz was estimated in \cite{Lu2023} to be ($64^{+4}_{-8} \, \mu \rm{as}$) , based on the circular ring fitting with the optimal set of \texttt{SMILI} images and visibility domain model fitting. Furthermore, they found that the thick ring is preferred over a thin ring using image analysis and model fitting. 


In this article, the posterior distribution of the M87 ring diameter, width, and ellipticity is estimated using the \texttt{variational image domain analysis} (\texttt{VIDA}) software \citep{VIDA}. To validate the robustness of the ring structure, \texttt{resolve} provides more reliable error estimates than the previous report since the uncertainty is estimated from posterior sample images, and not from the selected top-set of images. \texttt{VIDA} is an image feature extraction tool that treats each image as a probability distribution and compares the image to the geometrical model image (template) by utilizing Kullback-Leiber (KL) or Bhattacharyya (Bh) divergences as the objective function. 
Using the corresponding template, the ring features of each posterior sample image are estimated. In this analysis, we used the \texttt{CosineRingwFloor} template, and the azimuthal intensity distribution is described by a cosine expansion.

\begin{figure}[t]
    \centering
    \includegraphics[width=9cm]
    {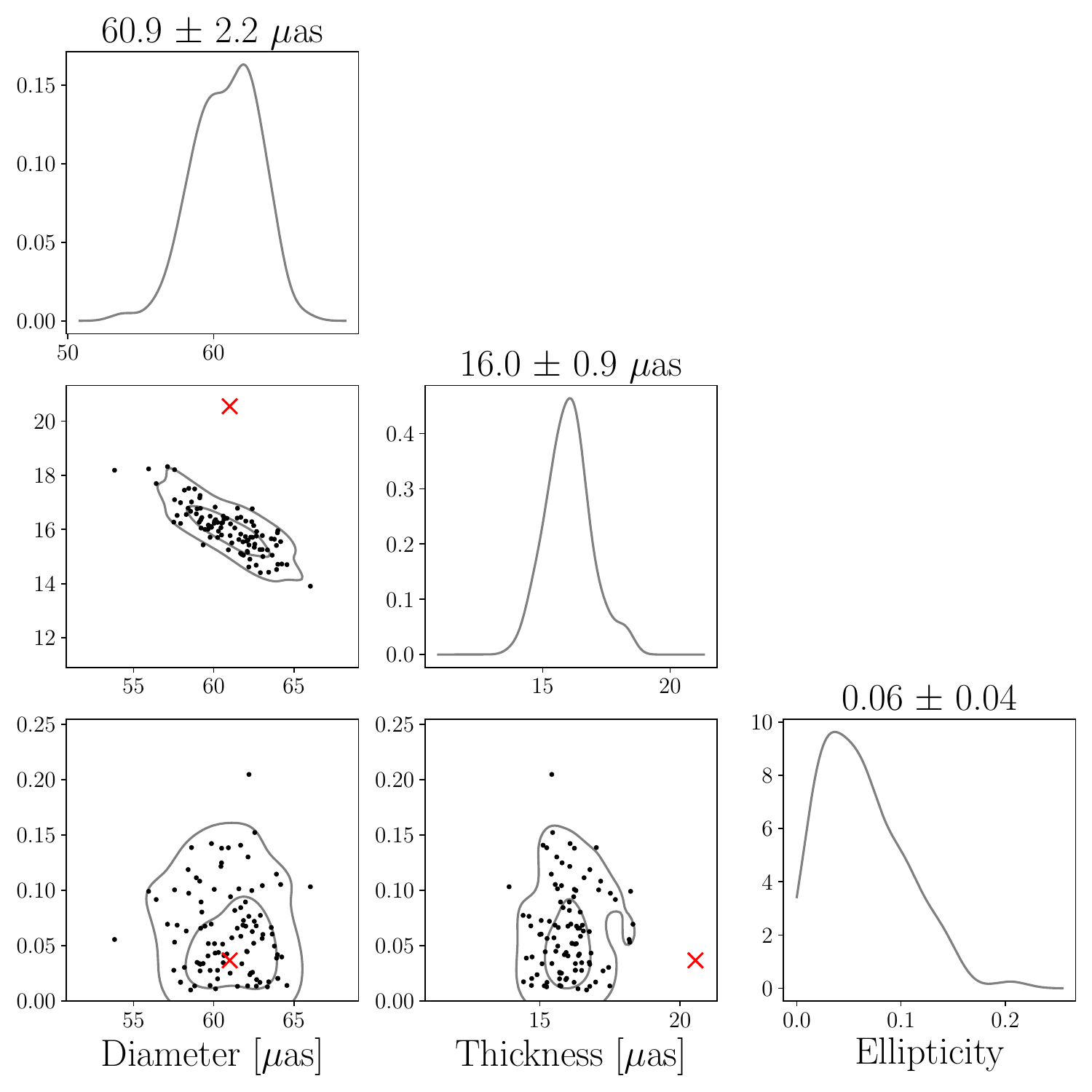}
    \caption{Posterior distribution of the M87 ring diameter, thickness, and ellipticity estimated from 100 \texttt{resolve} posterior sample images. Contours show 1$\sigma$ and 2$\sigma$ cumulative regions. The probability density function is obtained using Gaussian kernel density estimation. Each black dot marks the estimated ring parameter from the \texttt{resolve} posterior sample images. The red marks correspond to the estimation of the ring diameter and ring thickness obtained by the \texttt{DoG-HiT} reconstruction (Diameter : 61.0 $\mu \rm{as}$, Thickness : 20.6 $\mu \rm{as}$, Ellipticity : 0.04).
    }
    \label{fig:VIDA_posterior_plot}
\end{figure}

Figure \ref{fig:VIDA_posterior_plot} shows that the M87 ring diameter estimated from the \texttt{resolve} images is $60.9 \pm 2.2 \, \mu \rm{as}$ and is $61.0 \, \mu \rm{as}$ from the \texttt{DoG-HiT }image. The estimated ring diameter using \texttt{resolve} and \texttt{DoG-HiT} is within the errors of the estimation ($64^{+4}_{-8} \, \mu \rm{as}$) in \citep{Lu2023}. The width is $16.0 \pm 0.9 \, \mu \rm{as}$ using \texttt{resolve} and $20.6 \,\mu \rm{as}$ using \texttt{DoG-HiT}. The discrepancy in the ring width between \texttt{resolve} and \texttt{DoG-HiT} results from the sparse UV coverage beyond the first visibility null amplitude. The visibility domain model fitting \citep[see Fig. S10 in][] {Lu2023} shows that the UV coverage beyond the first null is crucial to determine the best fitting model. However, the thick ring and thin ring models show similar reduced $\chi^2$ due to the limited UV coverage beyond the first null. As a result, the ring width is not well constrained by the data.
Furthermore, we note that the ring diameter and width show anti-correlation, which is due to the finite resolution of the telescope array. The effective radius of the ring decreases with a larger ring width \citep[see Appendix G of][]{eht2019_L4_imaging}, which explains the dependence between the estimated diameter and width. This anti-correlation is also shown by \texttt{SMILI} reconstructions \citep[see Fig. S14 in][] {Lu2023}. The ellipticity of the ring, $\tau$, is defined as $\tau = 1 - b/a$, where $a$ is the semimajor axis lengths and $b$ is the semiminor axis lengths of the elliptical ring.
The $\tau$ is $0.06 \pm 0.04$ by \texttt{resolve} and $0.04$ by \texttt{DoG-HiT}, which means there is no significant ellipticity of the M87 ring. A detailed description of the \texttt{VIDA.jl} software can be found in \citet{VIDA}.

The observation by GMVA+ALMA M87  on 2018, April 14 was conducted a week before the EHT M87 observation (2018, April 21 and 25), which is shorter than the expected timescale for decorrelation of the emission pattern \citep{Georgiev2022}. 
The flux density of the compact region $200 \, \mu\rm{as} \, \times 200\, \mu$as) at 1\,mm is $0.5 \pm 0.1 $ Jy by \texttt{DIFMAP}, \texttt{eht-imaging}, and \texttt{SMILI} softwares (see Table~2 in \citealt{eht2024a}). We note that the compact region flux density constraints from EHT 2018 data were challenging due to the lack of short baseline coverage \citep{eht2024a}. The total flux density at 3\,mm is $0.57 \pm 0.03$\,Jy on mas scales, and the flux density of the compact region \footnote{Defined as the 200 $\mu$as (or about 50 $R_g$) field of view centered on the ring} at 3\,mm is $0.33 \pm 0.02$ Jy by \texttt{resolve}. 
\texttt{DoG-HiT} images estimate 0.43\,Jy in the compact field of view. As a result, the spectral index $\alpha$ of the M87 compact region is slightly positive $\alpha \sim 0.4$. This implies a mixed optical depth in the core, under a caveat that the emitting region at 3\,mm is larger than at 1\,mm, following the ring diameter analysis. The observed ratio of flux densities is reasonably consistent with the predictions of the numerical models, which typically indicate an inhomogeneous optical depth in the compact region \citep{EHT2019_P5, eht2021b, Palumbo2024}.


\begin{figure*}
    \centering
    \includegraphics[width=\textwidth]{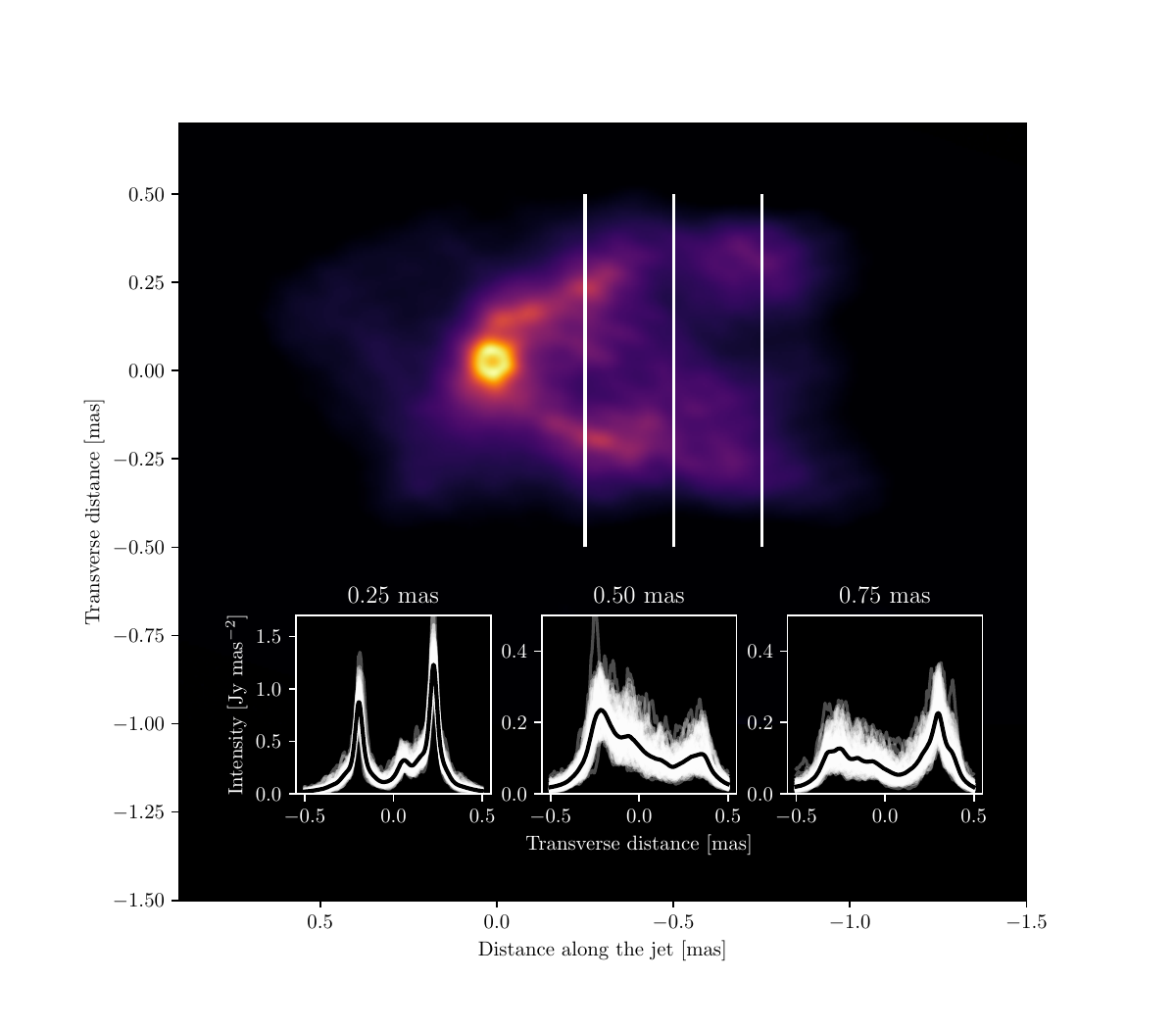}
    \caption{Galaxy M87 jet transverse profiles and intensity map obtained by \texttt{resolve}. Intensity in the map is represented by false color according to the color bar used in Fig. \ref{fig:summary} in logarithmic scale. The image is rotated $18^{\circ}$ clockwise. 
    Intensity plots at the bottom of the figure show flux density profiles of the jet at 0.25, 0.5 and 0.75\,mas from the phase center. Each vertical line corresponds to a location where profiles were extracted.}
    \label{fig:M87_jet_transverse_profile}
\end{figure*}

\subsection{Extended jet emission of M87}
Our reconstructions of the limb-brightened M87 jet structure broadly agree with the one described in \citet{Lu2023}: we see an edge-brightened jet anchored to the vicinity of the ring-like feature. One peculiar feature in the image reported by \citet{Lu2023} is the presence of a bright spine along the jet axis. This central ridge line may be related to the triple-helix structure in the jet of M87 that has been observed at larger scales \citep{Nikonov2023}. 

However, it is questionable whether this feature of the image represents a real structure on-sky, or appears as a consequence of imaging artifacts. Particularly, it has been recently demonstrated that \texttt{CLEAN} deconvolution errors are prone to produce inner ridge lines for edge-brightened jet configurations \citep{Pashchenko2023}. In fact, the central spine is less prominent in the \texttt{DoG-HiT} and \texttt{resolve} reconstructions.

\texttt{DoG-HiT} and \texttt{resolve} allow the robustness of the central spine to be quantified from two independent perspectives: by calibration-independent imaging with a minimal human bias in \texttt{DoG-HiT}, and by uncertainty estimation in \texttt{resolve}. For \texttt{DoG-HiT}, the small scale structure and the large scale (diffuse) structure are represented by different wavelets, which are ultimately expressed by the multiresolution support. This allows us to estimate the robustness of the central ridge by a jackknife test, that is, we cut the diffuse emission at the location of the central spine and recalculate the fit statistics to the (calibration-independent) closure quantities. To this end, we applied the following strategy. We flagged long baselines of more than $2\,G\lambda$ (to focus the analysis on the diffuse emission), then we fitted the closure phases and closure amplitudes with \texttt{DoG-HiT,} only varying coefficients in the multiresolution support, and calculated the updated fit statistics. Next, we masked out the diffuse, central spine from the multiresolution support, and refit the closure quantities. Finally, we compared the scoring with a central spine and without a central spine. 

This strategy resembles a standard strategy in VLBI, which is often applied in the discussion of the existence of a counter-jet. However, we note some key advantages of the strategy applied by us, compared to \texttt{CLEAN}. First, \texttt{DoG-HiT} directly fits closure quantities, hence the conclusions that we can draw are less dependent on the phase and amplitude self-calibration. Second, we perform the jackknife test on a multiscalar domain, which allows us to divide the emission more clearly into small and large scale structures. Finally, \texttt{DoG-HiT} directly fits a model to the data that is composed of diffuse and compact emission and does not need a final convolution with the beam, as opposed to the point source model of \texttt{CLEAN} \citep[e.g., see the discussion in][]{Mueller2023a}. Hence, we compare the fit quality of the approximated on-sky representation rather than an unphysical list of \texttt{CLEAN} components (which would question the interpretability of the $\chi^2$-statistics).

We obtain $\chi_{\textrm{cph}}^2 = 1.104$, $\chi_{\textrm{cla}}^2=1.494$ when fitting the data with a central spine, and $\chi_{\textrm{cph}}^2=1.061$, $\chi_{\textrm{cla}}^2=1.592$ when fitting without a central spine. Neither mode is strongly favored. From this study, we cannot report the conclusive detection of a central spine in the image.

An alternative perspective on the robustness of image features is offered by the built-in uncertainty quantification in \texttt{resolve}. In Fig. \ref{fig:M87_jet_transverse_profile}, the transverse flux intensity profiles of the M87 jet emission are depicted. The mean and standard deviation of the transverse jet profile can be obtained from 100 posterior sample images by \texttt{resolve}. The intensity profile of the jet at $0.25, 0.5, 0.75 \, \rm{mas}$ shows that the edge-brightened structure (two peaks) is prominent, however a significant central spine structure is not seen. In Fig. \ref{fig:M87_jet_transverse_profile}, the standard deviation of the pixel fluxes at the central spine is not particularly higher than the limb-brightened feature. In Fig. \ref{fig:summary}, the \texttt{DoG-HiT} image shows a central spine, which however is fainter than that of the \texttt{CLEAN} reconstruction. Therefore, the central spine in the images obtained by \texttt{CLEAN} in \citet{Lu2023} may be a consequence of \texttt{CLEAN} artifacts resulting from the \texttt{CLEAN} windows and sparsity-promoting \texttt{CLEAN} sky model. Further observations with additional short baseline antennas are required to conclude the detection of the central spine. The edge-brightened morphology that we recover is both consistent with the earlier observations at 86\,GHz \citep{JYKim2018} and well-motivated theoretically \citep[e.g., ][]{Yang2024}. The counter jet is not detected consistently in the three images, it is therefore not discussed further in this work.

\subsection{Substructure of the M87 ring}

A feature that appears consistent across all four different imaging algorithms \texttt{(SMILI}, \texttt{CLEAN}, \texttt{resolve}, and \texttt{DoG-HiT)} are the two bright blobs in the ring. While images are recovered with various resolutions, the recovered ring emission always shows the double structure within the ring toward the top and the bottom, at consistent positions across different imaging methods. We note that the limb-brightened structures that are recovered by \texttt{resolve} close to the ring seem to connect to the ring exactly at the positions where the brighter blobs within the ring occur consistently for all four reconstructions. Hence, it may be natural to interpret the double pattern in the ring as a physical phenomenon. In this subsection we present some discussion on how real these features may be. 

First, it is noticeable that these brighter regions in the ring form a double structure that is point-symmetric to the center of the ring. That supports the interpretation of these structures as an imaging artifact, especially due to the sparse coverage at long baselines. In fact, \citet{Lu2023} have tested \texttt{SMILI} and \texttt{CLEAN} reconstructions on synthetic data and found that similar structures are artificially introduced by the imaging procedure (compare Sect. 4 and particularly Fig. S8 in \citet{Lu2023}).

Multiple well-understood artifacts may cause such a double structure. Here, we discuss the three most natural scenarios. First, it could be caused by specific choices of the regularization assumption inherent to the respective imaging algorithm. Second, the structure could be introduced artificially by residual gain effects. Finally, the structure may be described by a residual side-lobe structure, which would make it essentially a consequence of the sparsity of the  UV coverage. In what follows, we will discuss each of these concerns individually.

Four imaging methods that were utilized here approach the image reconstruction from four vastly different perspectives: \texttt{CLEAN} recovers the structure in an inverse modeling framework essentially processing a sparsity-promoting regularization approach (sky brightness distribution is represented by a collection of delta components) \citep{Lannes1997}, \texttt{SMILI} approaches the image by a weighted sum of multiple handcrafted data and regularization terms in a RML framework \citep{Akiyama2017}; \texttt{DoG-HiT} processes multiscale functions in the context of compressive sensing \citep{Mueller2022}; and \texttt{resolve} estimates the posterior distribution of the image and gains from the prior model encoding source and instrument information and likelihood \citep[i.e., the data;][]{JSKim2024}. We note that the prior in Bayesian imaging can be interpreted as regularizers in RML methods, and vice versa \citep{JSKim2024}. The multiscale functions in \texttt{DoG-HiT} and the Gaussian process prior in \texttt{resolve} are flexible and do not ask for the double structure in the ring as prior knowledge.
The fact that all four independent methods that use a variety of prior information (regularization) lead to a similar structure challenges the interpretation of the double structure as an artifact from the assumptions and prior information applied by the imaging procedure.

The structure may be a consequence of unsolved gain residuals. In fact, it is a possible issue that the alternating self-calibration and cleaning procedure produce "phantom" structures point symmetric to the origin that has been reported in practice for a long time. We note, however, that the double feature also appears in the \texttt{DoG-HiT} reconstruction that are independent from gain corruptions (closure-only imaging), which makes a potential cause by the calibration of the phases less likely. This claim is supported by the \texttt{resolve} reconstruction which solves for the self-calibration with imaging simultaneously in a probabilistic setup. 

\begin{figure}
    \centering
    \includegraphics[width=9cm]{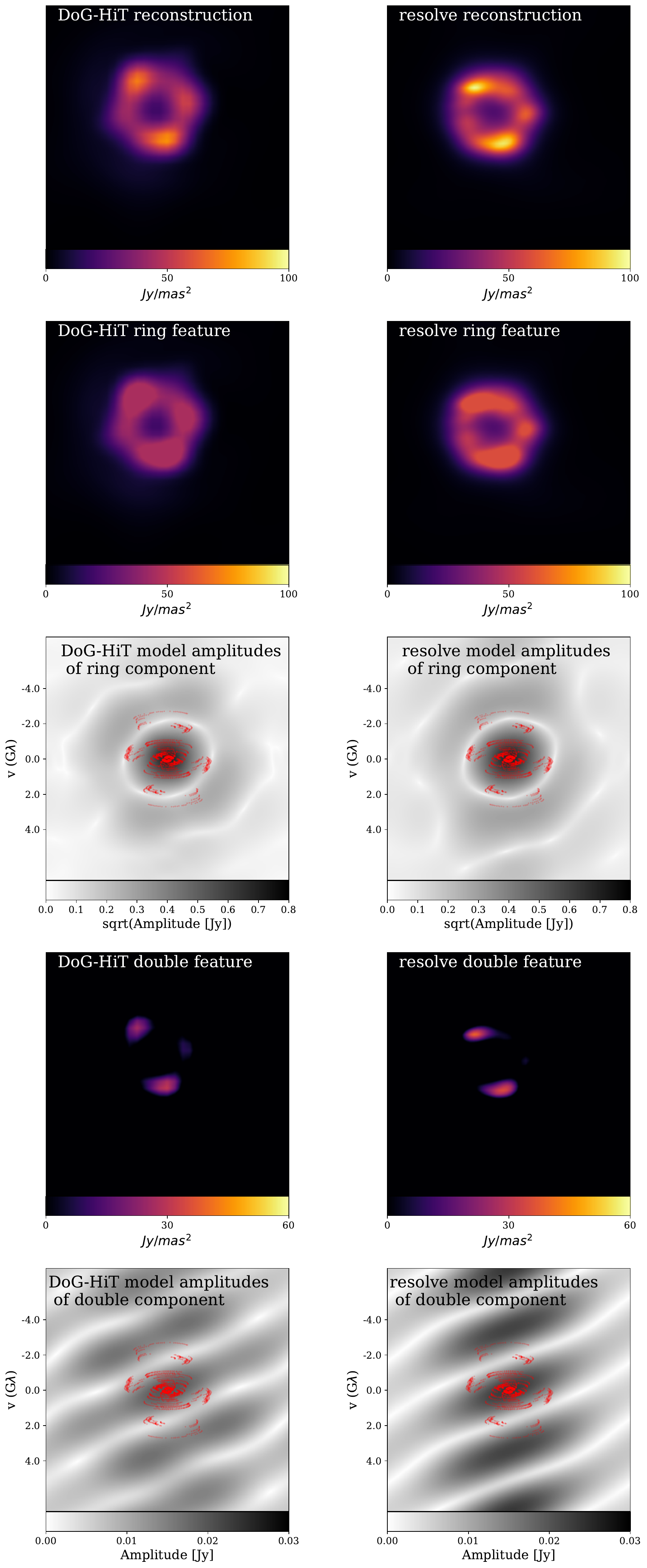}
    \caption{Reconstructions of the central feature by \texttt{DoG-HiT} (left panels) and \texttt{resolve} (right panels). Top panels: Reconstructions of the compact emission. Top-middle panels: The central ring feature. Middle panels: The recovered amplitudes of the central feature, with the UV coverage overplotted (red points). Middle-bottom panels: The central ring feature, cut at 60$\%$ of the respective peak brightness, which showcases the double blob pattern on top of the ring. Bottom panels: The amplitudes and the UV coverage of the double pattern alone.}
    \label{fig:uv_plot}
\end{figure}

Finally, we ask whether the double structure could be introduced by the sparsity of UV coverage. That is a possibility that  can never be eradicated completely, simply because the observation fundamentally misses relevant visibilities through the gaps in the UV coverage. Consequently, there is missing information. In fact, the Fourier domain representation of a double source is a fringe pattern (compare Fig. \ref{fig:uv_plot}), which is exactly the kind of artifact that not fully cleaned residuals may introduce. This issue has been identified through synthetic data tests by \citet{Lu2023}. They generated synthetic data sets with artificial baselines from symmetric ring ground truth images to understand the importance of the long west-east baselines. \texttt{CLEAN} and \texttt{SMILI} reconstructions from the synthetic data with additional baselines were able to reconstruct the ground truth. From the synthetic data with original UV coverage, \texttt{SMILI} reconstructions could recover symmetric ring ground truth images reasonably well, but \texttt{CLEAN} reconstructions tended to generate asymmetric ring structures. The occurrence of the double structure may be related to the lack of long west-east baselines. Nevertheless, we argue that the recovered double structure is represented by the measured visibilities. In Fig. \ref{fig:uv_plot}, we show the central ring image (top panels) and the amplitudes of the full Fourier transform of the reconstructions (middle panels) for \texttt{DoG-HiT} (left panels) and \texttt{resolve} (right panels). A ring feature is identified in the Fourier domain by a first zero that is clearly covered by observations \citep[compare, e.g., to the model fitting discussions in][]{Lu2023}. Moreover, the visibility domain representations of \texttt{DoG-HiT} and \texttt{resolve} show the ellipticity of the null visibility amplitude points in Fig. \ref{fig:uv_plot} (middle panel). We note that the location of the null visibility points in the UV domain is elliptical, with an elongation in the direction of the jet. This may result in bright spots in the image aligned perpendicularly to the jet, related to the stretch of the image domain ring. The data from multiple antennas (EF, KP, OV, PV, and YS) at the longest west-east baselines imply that the ellipticity of the null visibility points originates from the data.  

To analyze the UV-coverage sparsity that corresponds to the M87 ring substructure, we subtract a uniform ring feature from the images (defined by 60$\%$ of the respective emission peak) to extract the double feature on top of the ring. The resulting double patterns are depicted in the bottom panels of Fig. \ref{fig:uv_plot}, and their respective amplitudes in the Fourier domain in the last row. The observed UV points are overplotted with red crosses. Observations span the main fringe and the first side-lobe uniformly, the fringes are not produced exclusively in the gaps in the UV coverage.
These findings, as well as the striking similarity between multiple imaging approaches, constitute a convincing argument for the physical nature of bright spots along the ring. However, a definitive answer  cannot be given with the quality of the existing data set, particularly in the presence of the synthetic data tests performed by \citet{Lu2023}, and follow-up observations are needed. 

If the emission that forms the ring image is dominated by the accretion disk, elongation in the direction perpendicular to the ring may be a simple geometric effect for the inclined observer. While the 230\,GHz ring image is strongly asymmetric with respect to the jet axis \citep{eht2024a}, our 86\,GHz reconstructions exhibit a high degree of symmetry. In numerical general relativistic magnetohydrodynamic (GRMHD) simulations of accretion within the EHT library \citep{Dhruv2024}, consistent behavior appears for some retrograde accretion models (negative black hole spins). In that case, 230\,GHz image asymmetry is driven primarily by the spin effects \citep{EHT2019_P5}, which are dominant very near the event horizon. In the 86\,GHz image, which is formed by a more extended emission, these effects are balanced by the Doppler boost that enhances brightness on the opposite side of the black hole, which results in a ring structure elongated perpendicularly to the jet axis, with bright spots on both sides of the central brightness depression.



\section{Conclusions}
\label{Chap4}
\citet{Lu2023} present observations of the core and jet in M87 observed with the GMVA+ALMA at 86\,GHz. The image contains a ring-like feature that looks similar to the one reported by the EHT \citep{EHT2019_M87_paper1, eht2024a}, but with an approximately 1.5 times larger ring diameter resulting from the null visibility points and phase jump at 2.3 $G\lambda$. For the first time, this ring feature is connected to an innermost jet in the same image, possibly providing constraints on the launching mechanism of the jet. Furthermore, the recovered image contains several fainter features that may be of great importance for a scientific interpretation, especially in the context of recent works on the large scale jet structure in M87 \citep{JYKim2018, Cui2023, Nikonov2023, JYKim2023}.

To validate the results reported in \citet{Lu2023}, we apply two more imaging algorithms specially designed to study the robustness of the recovered features: by Bayesian self-calibration and imaging with \texttt{resolve}, and by closure-only imaging with \texttt{DoG-HiT}. The distinctive features of \texttt{resolve} and \texttt{Dog-HiT}, namely the probabilistic approach and the multiscale wavelet-based deconvolution algorithm, allow us to quantify the robustness of the recovered M87 ring and extended jet emission. We obtained the posterior distribution of the ring diameter, width, and ellipticity by analysis of the \texttt{resolve} posterior sample images. We confirm the M87 ring-like structure at 86\,GHz with a diameter of $60.9 \pm 2.2 \, \mu \rm{as}$, a thickness of $16.0 \pm 0.9 \, \mu \rm{as}$, and an ellipticity of $0.06 \pm 0.04$ by \texttt{resolve,} and a diameter of $61.0 \, \mu \rm{as}$, a thickness of $20.6 \, \mu \rm{as}$, and an ellipticity of $0.04$ by \texttt{DoG-HiT}. The estimated ring diameter is consistent with the estimate ($64^{+4}_{-8} \, \mu \rm{as}$) in \citet{Lu2023}.  

Furthermore, image reconstructions using \texttt{resolve} and \texttt{DoG-HiT} show that the ring is embedded in a strongly edge-brightened large scale jet structure, which agrees with the findings reported in \citet{Lu2023}. Among the upper and lower arm of the edge-brightened jet, the \texttt{CLEAN} reconstruction presented in \citet{Lu2023} features a third, central spine that may be interpreted together with a triple-helix structure at larger scales \citep{Nikonov2023}. However, the central spine structure in the \texttt{resolve} and \texttt{DoG-HiT} reconstructions is less prominent compared to the \texttt{CLEAN} reconstruction. Our analysis shows that this central spine is neither necessary to fit the data, nor supported by the uncertainty quantification by \texttt{resolve} in the image domain. The validation by two independent imaging methods implies that the central spine is fainter than previously reported.

Finally, all utilized imaging algorithms coincide on the same substructure in the ring consisting of two bright spots to the north and the south, co-located with inner anchor points on the edge-brightened jet to the horizon scale. Stemming from a coincidence of this phenomenon across a variety of imaging algorithms, and its representation in the Fourier domain, we argue that is more likely that this substructure of the ring results from the data, and not an imaging or self-calibration artifact, although an artifact that originates from sparse UV coverage is not ruled out. The potential physical origin of these structures is unclear. 
If real, they might be transient emission structures during the observational period, however, their alignment perpendicular to the jet suggests otherwise. They might also be permanent structures potentially linked to the disk-jet transition, as their location within the disk seems to coincide with the position the jet edges point to. Finally, they may be a result of an interplay of Doppler boost and black hole spin effects, as seen in numerical simulations of retrograde accretion. Investigating the size and asymmetry of the 86\,GHz emission ring in the EHT M87 simulation library could be a fruitful path toward understanding the detailed physics responsible for the morphology of our image.

\begin{acknowledgements}

      We thank Jae-Young Kim for helpful suggestions and feedback on drafts of the manuscript, Jack Livingston and Thomas Krichbaum for comments and informative discussions. J. K. and A. N. received financial support for this research from the International Max Planck Research School (IMPRS) for Astronomy and Astrophysics at the Universities of Bonn and Cologne. This work was supported by the M2FINDERS project funded by the European Research Council (ERC) under the European Union's Horizon 2020 Research and Innovation Programme (Grant Agreement No. 101018682). R.-S.L. is supported by the National Science Fund for Distinguished Young Scholars of China (Grant No. 12325302) and the Shanghai Pilot Program for Basic Research, CAS, Shanghai Branch (JCYJ-SHFY-2021-013). M. W. is supported by a Ramón y Cajal grant RYC2023-042988-I from the Spanish Ministry of Science and Innovation. This research has made use of data obtained with the Global Millimeter VLBI Array (GMVA), which consists of telescopes operated by the MPIfR, IRAM, Onsala, Metsahovi, Yebes, the Korean VLBI Network, the Greenland Telescope, the Green Bank Observatory and the Very Long Baseline Array (VLBA). The VLBA and the GBT are facilities of the National Science Foundation operated under cooperative agreement by Associated Universities, Inc. The Greenland Telescope (GLT) is operated by the Academia Sinica Institute of Astronomy and Astrophysics (ASIAA) and the Smithsonian Astrophysical Observatory (SAO). The data were correlated at the correlator of the MPIfR in Bonn, Germany.
\end{acknowledgements}

\bibliographystyle{aa}
\bibliography{reference}

\begin{thebibliography}{50}
\expandafter\ifx\csname natexlab\endcsname\relax\def\natexlab#1{#1}\fi

\bibitem[{{Akiyama} {et~al.}(2017){Akiyama}, {Kuramochi}, {Ikeda}, {Fish}, {Tazaki}, {Honma}, {Doeleman}, {Broderick}, {Dexter}, {Mo{\'s}cibrodzka}, {Bouman}, {Chael}, \& {Zaizen}}]{Akiyama2017}
{Akiyama}, K., {Kuramochi}, K., {Ikeda}, S., {et~al.} 2017, \apj, 838, 1

\bibitem[{{Arras} {et~al.}(2021){Arras}, {Bester}, {Perley}, {Leike}, {Smirnov}, {Westermann}, \& {En{\ss}lin}}]{Arras_2021_CygA}
{Arras}, P., {Bester}, H.~L., {Perley}, R.~A., {et~al.} 2021, \aap, 646, A84

\bibitem[{{Arras} {et~al.}(2022){Arras}, {Frank}, {Haim}, {Knollm{\"u}ller}, {Leike}, {Reinecke}, \& {En{\ss}lin}}]{Arras2022}
{Arras}, P., {Frank}, P., {Haim}, P., {et~al.} 2022, Nature Astronomy, 6, 259

\bibitem[{{Asada} {et~al.}(2016){Asada}, {Nakamura}, \& {Pu}}]{2016ApJ...833...56A}
{Asada}, K., {Nakamura}, M., \& {Pu}, H.-Y. 2016, \apj, 833, 56

\bibitem[{{Blackburn} {et~al.}(2020){Blackburn}, {Pesce}, {Johnson}, {Wielgus}, {Chael}, {Christian}, \& {Doeleman}}]{Blackburn2020}
{Blackburn}, L., {Pesce}, D.~W., {Johnson}, M.~D., {et~al.} 2020, \apj, 894, 31

\bibitem[{{Blei} {et~al.}(2016){Blei}, {Kucukelbir}, \& {McAuliffe}}]{Blei2016_VI_review}
{Blei}, D.~M., {Kucukelbir}, A., \& {McAuliffe}, J.~D. 2016, arXiv e-prints, arXiv:1601.00670

\bibitem[{{Cappellari} {et~al.}(2011){Cappellari}, {Emsellem}, {Krajnovi{\'c}}, {McDermid}, {Scott}, {Verdoes Kleijn}, {Young}, {Alatalo}, {Bacon}, {Blitz}, {Bois}, {Bournaud}, {Bureau}, {Davies}, {Davis}, {de Zeeuw}, {Duc}, {Khochfar}, {Kuntschner}, {Lablanche}, {Morganti}, {Naab}, {Oosterloo}, {Sarzi}, {Serra}, \& {Weijmans}}]{2011MNRAS.413..813C}
{Cappellari}, M., {Emsellem}, E., {Krajnovi{\'c}}, D., {et~al.} 2011, \mnras, 413, 813

\bibitem[{{Chael} {et~al.}(2018){Chael}, {Johnson}, {Bouman}, {Blackburn}, {Akiyama}, \& {Narayan}}]{Chael2018}
{Chael}, A.~A., {Johnson}, M.~D., {Bouman}, K.~L., {et~al.} 2018, \apj, 857, 23

\bibitem[{{Chael} {et~al.}(2016){Chael}, {Johnson}, {Narayan}, {Doeleman}, {Wardle}, \& {Bouman}}]{Chael2016}
{Chael}, A.~A., {Johnson}, M.~D., {Narayan}, R., {et~al.} 2016, \apj, 829, 11

\bibitem[{{Clark}(1980)}]{1980A&A....89..377C}
{Clark}, B.~G. 1980, \aap, 89, 377

\bibitem[{{Cui} {et~al.}(2023){Cui}, {Hada}, {Kawashima}, {Kino}, {Lin}, {Mizuno}, {Ro}, {Honma}, {Yi}, {Yu}, {Park}, {Jiang}, {Shen}, {Kravchenko}, {Algaba}, {Cheng}, {Cho}, {Giovannini}, {Giroletti}, {Jung}, {Lu}, {Niinuma}, {Oh}, {Ohsuga}, {Sawada-Satoh}, {Sohn}, {Takahashi}, {Takamura}, {Tazaki}, {Trippe}, {Wajima}, {Akiyama}, {An}, {Asada}, {Buttaccio}, {Byun}, {Cui}, {Hagiwara}, {Hirota}, {Hodgson}, {Kawaguchi}, {Kim}, {Lee}, {Lee}, {Lee}, {Maccaferri}, {Melis}, {Melnikov}, {Migoni}, {Oh}, {Sugiyama}, {Wang}, {Zhang}, {Chen}, {Hwang}, {Jung}, {Kim}, {Kim}, {Kobayashi}, {Li}, {Li}, {Li}, {Liu}, {Liu}, {Liu}, {Oh}, {Oyama}, {Roh}, {Wang}, {Wang}, {Wang}, {Xia}, {Yan}, {Yeom}, {Yonekura}, {Yuan}, {Zhang}, {Zhao}, \& {Zhong}}]{Cui2023}
{Cui}, Y., {Hada}, K., {Kawashima}, T., {et~al.} 2023, \nat, 621, 711

\bibitem[{{Dhruv} {et~al.}(2024){Dhruv}, {Prather}, {Wong}, \& {Gammie}}]{Dhruv2024}
{Dhruv}, V., {Prather}, B., {Wong}, G., \& {Gammie}, C.~F. 2024, arXiv e-prints, arXiv:2411.12647

\bibitem[{{Event Horizon Telescope Collaboration} {et~al.}(2024){Event Horizon Telescope Collaboration}, {Akiyama}, {Alberdi}, {Alef}, {Algaba}, {Anantua}, {Asada}, {Azulay}, {Bach}, {Baczko}, {Ball}, {Balokovi{\'c}}, {Bandyopadhyay}, {Barrett}, {Baub{\"o}ck}, {Benson}, {Bintley}, {Blackburn}, {Blundell}, {Bouman}, {Bower}, {Boyce}, {Bremer}, {Brissenden}, {Britzen}, {Broderick}, {Broguiere}, {Bronzwaer}, {Bustamante}, {Carlstrom}, {Chael}, {Chan}, {Chang}, {Chatterjee}, {Chatterjee}, {Chen}, {Chen}, {Cheng}, {Cho}, {Christian}, {Conroy}, {Conway}, {Crawford}, {Crew}, {Cruz-Osorio}, {Cui}, {Dahale}, {Davelaar}, {De Laurentis}, {Deane}, {Dempsey}, {Desvignes}, {Dexter}, {Dhruv}, {Dihingia}, {Doeleman}, {Dzib}, {Eatough}, {Emami}, {Falcke}, {Farah}, {Fish}, {Fomalont}, {Ford}, {Foschi}, {Fraga-Encinas}, {Freeman}, {Friberg}, {Fromm}, {Fuentes}, {Galison}, {Gammie}, {Garc{\'\i}a}, {Gentaz}, {Georgiev}, {Goddi}, {Gold}, {G{\'o}mez-Ruiz}, {G{\'o}mez}, {Gu}, {Gurwell}, {Hada}, {Haggard}, {Hesper}, {Heumann}, {Ho},
  {Ho}, {Honma}, {Huang}, {Huang}, {Hughes}, {Ikeda}, {Violette Impellizzeri}, {Inoue}, {Issaoun}, {James}, {Jannuzi}, {Janssen}, {Jeter}, {Jiang}, {Jim{\'e}nez-Rosales}, {Johnson}, {Jorstad}, {Jones}, {Joshi}, {Jung}, {Karuppusamy}, {Kawashima}, {Keating}, {Kettenis}, {Kim}, {Kim}, {Kim}, {Kim}, {Kino}, {Koay}, {Kocherlakota}, {Kofuji}, {Koch}, {Koyama}, {Kramer}, {Kramer}, {Kramer}, {Krichbaum}, {Kuo}, {La Bella}, {Lee}, {Levis}, {Li}, {Lico}, {Lindahl}, {Lindqvist}, {Lisakov}, {Liu}, {Liu}, {Liuzzo}, {Lo}, {Lobanov}, {Loinard}, {Lonsdale}, {Lowitz}, {Lu}, {MacDonald}, {Mao}, {Marchili}, {Markoff}, {Marrone}, {Marscher}, {Mart{\'\i}-Vidal}, {Matsushita}, {Matthews}, {Medeiros}, {Menten}, {Mizuno}, {Mizuno}, {Montgomery}, {Moran}, {Moriyama}, {Moscibrodzka}, {Mulaudzi}, {M{\"u}ller}, {M{\"u}ller}, {Mus}, {Musoke}, {Myserlis}, {Nagai}, {Nagar}, {Nakamura}, {Narayanan}, {Natarajan}, {Nathanail}, {Fuentes}, {Neilsen}, {Ni}, {Nowak}, {Oh}, {Okino}, {Olivares}, {Oyama}, {{\"O}zel}, {Palumbo}, {Paraschos}, {Park},
  {Parsons}, {Patel}, {Pen}, {Pesce}, {Pi{\'e}tu}, {PopStefanija}, {Porth}, {Prather}, {Psaltis}, {Pu}, {Ramakrishnan}, {Rao}, {Rawlings}, {Raymond}, {Rezzolla}, {Ricarte}, {Ripperda}, {Roelofs}, {Romero-Ca{\~n}izales}, {Ros}, {Roshanineshat}, {Rottmann}, {Roy}, {Ruiz}, {Ruszczyk}, {Rygl}, {S{\'a}nchez}, {S{\'a}nchez-Arg{\"u}elles}, {S{\'a}nchez-Portal}, {Sasada}, {Satapathy}, {Savolainen}, {Schloerb}, {Schonfeld}, {Schuster}, {Shao}, {Shen}, {Small}, {Sohn}, {SooHoo}, {Salas}, {Souccar}, {Stanway}, {Sun}, {Tazaki}, {Tetarenko}, {Tiede}, {Tilanus}, {Titus}, {Toma}, {Torne}, {Toscano}, {Traianou}, {Trent}, {Trippe}, {Turk}, {van Bemmel}, {van Langevelde}, {van Rossum}, {Vos}, {Wagner}, {Ward-Thompson}, {Wardle}, {Washington}, {Weintroub}, {Wharton}, {Wielgus}, {Wiik}, {Witzel}, {Wondrak}, {Wong}, {Wu}, {Yadlapalli}, {Yamaguchi}, {Yfantis}, {Yoon}, {Young}, {Younsi}, {Yu}, {Yuan}, {Yuan}, {Anton Zensus}, {Zhang}, {Zhao}, {Zhao}, {Allardi}, {Chang}, {Chang}, {Chang}, {Chen}, {Chilson}, {Faber}, {Gale}, {Han},
  {Han}, {Hasegawa}, {Hern{\'a}ndez-Rebollar}, {Huang}, {Jiang}, {Jinchi}, {Kimura}, {Kubo}, {Li}, {Lin}, {Liu}, {Liu}, {Lu}, {Martin-Cocher}, {Meyer-Zhao}, {Monta{\~n}a}, {Moraghan}, {Moreno-Nolasco}, {Nishioka}, {Norton}, {Nystrom}, {Ogawa}, {Oshiro}, {Pradel}, {Principe}, {Raffin}, {Rodr{\'\i}guez-Montoya}, {Shaw}, {Snow}, {Sridharan}, {Srinivasan}, {Wei}, \& {Yu}}]{eht2024a}
{Event Horizon Telescope Collaboration}, {Akiyama}, K., {Alberdi}, A., {et~al.} 2024, \aap, 681, A79

\bibitem[{{Event Horizon Telescope Collaboration} {et~al.}(2023){Event Horizon Telescope Collaboration}, {Akiyama}, {Alberdi}, {Alef}, {Algaba}, {Anantua}, {Asada}, {Azulay}, {Bach}, {Baczko}, {Ball}, {Balokovi{\'c}}, {Barrett}, {Baub{\"o}ck}, {Benson}, {Bintley}, {Blackburn}, {Blundell}, {Bouman}, {Bower}, {Boyce}, {Bremer}, {Brinkerink}, {Brissenden}, {Britzen}, {Broderick}, {Broguiere}, {Bronzwaer}, {Bustamante}, {Byun}, {Carlstrom}, {Ceccobello}, {Chael}, {Chan}, {Chang}, {Chatterjee}, {Chatterjee}, {Chen}, {Chen}, {Cheng}, {Cho}, {Christian}, {Conroy}, {Conway}, {Cordes}, {Crawford}, {Crew}, {Cruz-Osorio}, {Cui}, {Dahale}, {Davelaar}, {De Laurentis}, {Deane}, {Dempsey}, {Desvignes}, {Dexter}, {Dhruv}, {Doeleman}, {Dougal}, {Dzib}, {Eatough}, {Emami}, {Falcke}, {Farah}, {Fish}, {Fomalont}, {Ford}, {Foschi}, {Fraga-Encinas}, {Freeman}, {Friberg}, {Fromm}, {Fuentes}, {Galison}, {Gammie}, {Garc{\'\i}a}, {Gentaz}, {Georgiev}, {Goddi}, {Gold}, {G{\'o}mez-Ruiz}, {G{\'o}mez}, {Gu}, {Gurwell}, {Hada}, {Haggard},
  {Haworth}, {Hecht}, {Hesper}, {Heumann}, {Ho}, {Ho}, {Honma}, {Huang}, {Huang}, {Hughes}, {Ikeda}, {Impellizzeri}, {Inoue}, {Issaoun}, {James}, {Jannuzi}, {Janssen}, {Jeter}, {Jiang}, {Jim{\'e}nez-Rosales}, {Johnson}, {Jorstad}, {Joshi}, {Jung}, {Karami}, {Karuppusamy}, {Kawashima}, {Keating}, {Kettenis}, {Kim}, {Kim}, {Kim}, {Kim}, {Kino}, {Koay}, {Kocherlakota}, {Kofuji}, {Koch}, {Koyama}, {Kramer}, {Kramer}, {Kramer}, {Krichbaum}, {Kuo}, {La Bella}, {Lauer}, {Lee}, {Lee}, {Leung}, {Levis}, {Li}, {Lico}, {Lindahl}, {Lindqvist}, {Lisakov}, {Liu}, {Liu}, {Liuzzo}, {Lo}, {Lobanov}, {Loinard}, {Lonsdale}, {Lowitz}, {Lu}, {MacDonald}, {Mao}, {Marchili}, {Markoff}, {Marrone}, {Marscher}, {Mart{\'\i}-Vidal}, {Matsushita}, {Matthews}, {Medeiros}, {Menten}, {Michalik}, {Mizuno}, {Mizuno}, {Moran}, {Moriyama}, {Moscibrodzka}, {Mulaudzi}, {M{\"u}ller}, {M{\"u}ller}, {Mus}, {Musoke}, {Myserlis}, {Nadolski}, {Nagai}, {Nagar}, {Nakamura}, {Narayan}, {Narayanan}, {Natarajan}, {Nathanail}, {Fuentes}, {Neilsen}, {Neri},
  {Ni}, {Noutsos}, {Nowak}, {Oh}, {Okino}, {Olivares}, {Ortiz-Le{\'o}n}, {Oyama}, {{\"O}zel}, {Palumbo}, {Paraschos}, {Park}, {Parsons}, {Patel}, {Pen}, {Pesce}, {Pi{\'e}tu}, {Plambeck}, {PopStefanija}, {Porth}, {P{\"o}tzl}, {Prather}, {Preciado-L{\'o}pez}, {Psaltis}, {Pu}, {Ramakrishnan}, {Rao}, {Rawlings}, {Raymond}, {Rezzolla}, {Ricarte}, {Ripperda}, {Roelofs}, {Rogers}, {Romero-Ca{\~n}izales}, {Ros}, {Roshanineshat}, {Rottmann}, {Roy}, {Ruiz}, {Ruszczyk}, {Rygl}, {S{\'a}nchez}, {S{\'a}nchez-Arg{\"u}elles}, {S{\'a}nchez-Portal}, {Sasada}, {Satapathy}, {Savolainen}, {Schloerb}, {Schonfeld}, {Schuster}, {Shao}, {Shen}, {Small}, {Sohn}, {SooHoo}, {Sosapanta Salas}, {Souccar}, {Sun}, {Tazaki}, {Tetarenko}, {Tiede}, {Tilanus}, {Titus}, {Torne}, {Toscano}, {Traianou}, {Trent}, {Trippe}, {Turk}, {van Bemmel}, {van Langevelde}, {van Rossum}, {Vos}, {Wagner}, {Ward-Thompson}, {Wardle}, {Washington}, {Weintroub}, {Wharton}, {Wielgus}, {Wiik}, {Witzel}, {Wondrak}, {Wong}, {Wu}, {Yadlapalli}, {Yamaguchi}, {Yfantis},
  {Yoon}, {Young}, {Young}, {Younsi}, {Yu}, {Yuan}, {Yuan}, {Zensus}, {Zhang}, {Zhao}, \& {Zhao}}]{eht2023circ}
{Event Horizon Telescope Collaboration}, {Akiyama}, K., {Alberdi}, A., {et~al.} 2023, \apjl, 957, L20

\bibitem[{{Event Horizon Telescope Collaboration} {et~al.}(2022){Event Horizon Telescope Collaboration}, {Akiyama}, {Alberdi}, {Alef}, {Algaba}, {Anantua}, {Asada}, {Azulay}, {Bach}, {Baczko}, {Ball}, {Balokovi{\'c}}, {Barrett}, {Baub{\"o}ck}, {Benson}, {Bintley}, {Blackburn}, {Blundell}, {Bouman}, {Bower}, {Boyce}, {Bremer}, {Brinkerink}, {Brissenden}, {Britzen}, {Broderick}, {Broguiere}, {Bronzwaer}, {Bustamante}, {Byun}, {Carlstrom}, {Ceccobello}, {Chael}, {Chan}, {Chatterjee}, {Chatterjee}, {Chen}, {Chen}, {Cheng}, {Cho}, {Christian}, {Conroy}, {Conway}, {Cordes}, {Crawford}, {Crew}, {Cruz-Osorio}, {Cui}, {Davelaar}, {De Laurentis}, {Deane}, {Dempsey}, {Desvignes}, {Dexter}, {Dhruv}, {Doeleman}, {Dougal}, {Dzib}, {Eatough}, {Emami}, {Falcke}, {Farah}, {Fish}, {Fomalont}, {Ford}, {Fraga-Encinas}, {Freeman}, {Friberg}, {Fromm}, {Fuentes}, {Galison}, {Gammie}, {Garc{\'\i}a}, {Gentaz}, {Georgiev}, {Goddi}, {Gold}, {G{\'o}mez-Ruiz}, {G{\'o}mez}, {Gu}, {Gurwell}, {Hada}, {Haggard}, {Haworth}, {Hecht}, {Hesper},
  {Heumann}, {Ho}, {Ho}, {Honma}, {Huang}, {Huang}, {Hughes}, {Ikeda}, {Impellizzeri}, {Inoue}, {Issaoun}, {James}, {Jannuzi}, {Janssen}, {Jeter}, {Jiang}, {Jim{\'e}nez-Rosales}, {Johnson}, {Jorstad}, {Joshi}, {Jung}, {Karami}, {Karuppusamy}, {Kawashima}, {Keating}, {Kettenis}, {Kim}, {Kim}, {Kim}, {Kim}, {Kino}, {Koay}, {Kocherlakota}, {Kofuji}, {Koch}, {Koyama}, {Kramer}, {Kramer}, {Krichbaum}, {Kuo}, {La Bella}, {Lauer}, {Lee}, {Lee}, {Leung}, {Levis}, {Li}, {Lico}, {Lindahl}, {Lindqvist}, {Lisakov}, {Liu}, {Liu}, {Liuzzo}, {Lo}, {Lobanov}, {Loinard}, {Lonsdale}, {Lu}, {Mao}, {Marchili}, {Markoff}, {Marrone}, {Marscher}, {Mart{\'\i}-Vidal}, {Matsushita}, {Matthews}, {Medeiros}, {Menten}, {Michalik}, {Mizuno}, {Mizuno}, {Moran}, {Moriyama}, {Moscibrodzka}, {M{\"u}ller}, {Mus}, {Musoke}, {Myserlis}, {Nadolski}, {Nagai}, {Nagar}, {Nakamura}, {Narayan}, {Narayanan}, {Natarajan}, {Nathanail}, {Fuentes}, {Neilsen}, {Neri}, {Ni}, {Noutsos}, {Nowak}, {Oh}, {Okino}, {Olivares}, {Ortiz-Le{\'o}n}, {Oyama},
  {{\"O}zel}, {Palumbo}, {Paraschos}, {Park}, {Parsons}, {Patel}, {Pen}, {Pesce}, {Pi{\'e}tu}, {Plambeck}, {PopStefanija}, {Porth}, {P{\"o}tzl}, {Prather}, {Preciado-L{\'o}pez}, {Psaltis}, {Pu}, {Ramakrishnan}, {Rao}, {Rawlings}, {Raymond}, {Rezzolla}, {Ricarte}, {Ripperda}, {Roelofs}, {Rogers}, {Ros}, {Romero-Ca{\~n}izales}, {Roshanineshat}, {Rottmann}, {Roy}, {Ruiz}, {Ruszczyk}, {Rygl}, {S{\'a}nchez}, {S{\'a}nchez-Arg{\"u}elles}, {S{\'a}nchez-Portal}, {Sasada}, {Satapathy}, {Savolainen}, {Schloerb}, {Schonfeld}, {Schuster}, {Shao}, {Shen}, {Small}, {Sohn}, {SooHoo}, {Souccar}, {Sun}, {Tazaki}, {Tetarenko}, {Tiede}, {Tilanus}, {Titus}, {Torne}, {Traianou}, {Trent}, {Trippe}, {Turk}, {van Bemmel}, {van Langevelde}, {van Rossum}, {Vos}, {Wagner}, {Ward-Thompson}, {Wardle}, {Weintroub}, {Wex}, {Wharton}, {Wielgus}, {Wiik}, {Witzel}, {Wondrak}, {Wong}, {Wu}, {Yamaguchi}, {Yoon}, {Young}, {Young}, {Younsi}, {Yuan}, {Yuan}, {Zensus}, {Zhang}, {Zhao}, {Zhao}, {Agurto}, {Allardi}, {Amestica}, {Araneda}, {Arriagada},
  {Berghuis}, {Bertarini}, {Berthold}, {Blanchard}, {Brown}, {C{\'a}rdenas}, {Cantzler}, {Caro}, {Castillo-Dom{\'\i}nguez}, {Chan}, {Chang}, {Chang}, {Chang}, {Chang}, {Chen}, {Chilson}, {Chuter}, {Ciechanowicz}, {Colin-Beltran}, {Coulson}, {Crowley}, {Degenaar}, {Dornbusch}, {Dur{\'a}n}, {Everett}, {Faber}, {Forster}, {Fuchs}, {Gale}, {Geertsema}, {Gonz{\'a}lez}, {Graham}, {Gueth}, {Halverson}, {Han}, {Han}, {Hasegawa}, {Hern{\'a}ndez-Rebollar}, {Herrera}, {Herrero-Illana}, {Heyminck}, {Hirota}, {Hoge}, {Hostler Schimpf}, {Howie}, {Huang}, {Jiang}, {Jinchi}, {John}, {Kimura}, {Klein}, {Kubo}, {Kuroda}, {Kwon}, {Lacasse}, {Laing}, {Leitch}, {Li}, {Liu}, {Liu}, {Lin}, {Lu}, {Mac-Auliffe}, {Martin-Cocher}, {Matulonis}, {Maute}, {Messias}, {Meyer-Zhao}, {Monta{\~n}a}, {Montenegro-Montes}, {Montgomerie}, {Moreno Nolasco}, {Muders}, {Nishioka}, {Norton}, {Nystrom}, {Ogawa}, {Olivares}, {Oshiro}, {P{\'e}rez-Beaupuits}, {Parra}, {Phillips}, {Poirier}, {Pradel}, {Qiu}, {Raffin}, {Rahlin}, {Ram{\'\i}rez}, {Ressler},
  {Reynolds}, {Rodr{\'\i}guez-Montoya}, {Saez-Madain}, {Santana}, {Shaw}, {Shirkey}, {Silva}, {Snow}, {Sousa}, {Sridharan}, {Stahm}, {Stark}, {Test}, {Torstensson}, {Venegas}, {Walther}, {Wei}, {White}, {Wieching}, {Wijnands}, {Wouterloot}, {Yu}, {Yu (于威)}, \& {Zeballos}}]{eht2022a}
{Event Horizon Telescope Collaboration}, {Akiyama}, K., {Alberdi}, A., {et~al.} 2022, \apjl, 930, L12

\bibitem[{{Event Horizon Telescope Collaboration} {et~al.}(2019{\natexlab{a}}){Event Horizon Telescope Collaboration}, {Akiyama}, {Alberdi}, {Alef}, {Asada}, {Azulay}, {Baczko}, {Ball}, {Balokovi{\'c}}, {Barrett}, {Bintley}, {Blackburn}, {Boland}, {Bouman}, {Bower}, {Bremer}, {Brinkerink}, {Brissenden}, {Britzen}, {Broderick}, {Broguiere}, {Bronzwaer}, {Byun}, {Carlstrom}, {Chael}, {Chan}, {Chatterjee}, {Chatterjee}, {Chen}, {Chen}, {Cho}, {Christian}, {Conway}, {Cordes}, {Crew}, {Cui}, {Davelaar}, {De Laurentis}, {Deane}, {Dempsey}, {Desvignes}, {Dexter}, {Doeleman}, {Eatough}, {Falcke}, {Fish}, {Fomalont}, {Fraga-Encinas}, {Freeman}, {Friberg}, {Fromm}, {G{\'o}mez}, {Galison}, {Gammie}, {Garc{\'\i}a}, {Gentaz}, {Georgiev}, {Goddi}, {Gold}, {Gu}, {Gurwell}, {Hada}, {Hecht}, {Hesper}, {Ho}, {Ho}, {Honma}, {Huang}, {Huang}, {Hughes}, {Ikeda}, {Inoue}, {Issaoun}, {James}, {Jannuzi}, {Janssen}, {Jeter}, {Jiang}, {Johnson}, {Jorstad}, {Jung}, {Karami}, {Karuppusamy}, {Kawashima}, {Keating}, {Kettenis}, {Kim},
  {Kim}, {Kim}, {Kino}, {Koay}, {Koch}, {Koyama}, {Kramer}, {Kramer}, {Krichbaum}, {Kuo}, {Lauer}, {Lee}, {Li}, {Li}, {Lindqvist}, {Liu}, {Liuzzo}, {Lo}, {Lobanov}, {Loinard}, {Lonsdale}, {Lu}, {MacDonald}, {Mao}, {Markoff}, {Marrone}, {Marscher}, {Mart{\'\i}-Vidal}, {Matsushita}, {Matthews}, {Medeiros}, {Menten}, {Mizuno}, {Mizuno}, {Moran}, {Moriyama}, {Moscibrodzka}, {M{\"u}ller}, {Nagai}, {Nagar}, {Nakamura}, {Narayan}, {Narayanan}, {Natarajan}, {Neri}, {Ni}, {Noutsos}, {Okino}, {Olivares}, {Ortiz-Le{\'o}n}, {Oyama}, {{\"O}zel}, {Palumbo}, {Patel}, {Pen}, {Pesce}, {Pi{\'e}tu}, {Plambeck}, {PopStefanija}, {Porth}, {Prather}, {Preciado-L{\'o}pez}, {Psaltis}, {Pu}, {Ramakrishnan}, {Rao}, {Rawlings}, {Raymond}, {Rezzolla}, {Ripperda}, {Roelofs}, {Rogers}, {Ros}, {Rose}, {Roshanineshat}, {Rottmann}, {Roy}, {Ruszczyk}, {Ryan}, {Rygl}, {S{\'a}nchez}, {S{\'a}nchez-Arguelles}, {Sasada}, {Savolainen}, {Schloerb}, {Schuster}, {Shao}, {Shen}, {Small}, {Sohn}, {SooHoo}, {Tazaki}, {Tiede}, {Tilanus}, {Titus}, {Toma},
  {Torne}, {Trent}, {Trippe}, {Tsuda}, {van Bemmel}, {van Langevelde}, {van Rossum}, {Wagner}, {Wardle}, {Weintroub}, {Wex}, {Wharton}, {Wielgus}, {Wong}, {Wu}, {Young}, {Young}, {Younsi}, {Yuan}, {Yuan}, {Zensus}, {Zhao}, {Zhao}, {Zhu}, {Algaba}, {Allardi}, {Amestica}, {Anczarski}, {Bach}, {Baganoff}, {Beaudoin}, {Benson}, {Berthold}, {Blanchard}, {Blundell}, {Bustamente}, {Cappallo}, {Castillo-Dom{\'\i}nguez}, {Chang}, {Chang}, {Chang}, {Chen}, {Chilson}, {Chuter}, {C{\'o}rdova Rosado}, {Coulson}, {Crawford}, {Crowley}, {David}, {Derome}, {Dexter}, {Dornbusch}, {Dudevoir}, {Dzib}, {Eckart}, {Eckert}, {Erickson}, {Everett}, {Faber}, {Farah}, {Fath}, {Folkers}, {Forbes}, {Freund}, {G{\'o}mez-Ruiz}, {Gale}, {Gao}, {Geertsema}, {Graham}, {Greer}, {Grosslein}, {Gueth}, {Haggard}, {Halverson}, {Han}, {Han}, {Hao}, {Hasegawa}, {Henning}, {Hern{\'a}ndez-G{\'o}mez}, {Herrero-Illana}, {Heyminck}, {Hirota}, {Hoge}, {Huang}, {Impellizzeri}, {Jiang}, {Kamble}, {Keisler}, {Kimura}, {Kono}, {Kubo}, {Kuroda}, {Lacasse},
  {Laing}, {Leitch}, {Li}, {Lin}, {Liu}, {Liu}, {Lu}, {Marson}, {Martin-Cocher}, {Massingill}, {Matulonis}, {McColl}, {McWhirter}, {Messias}, {Meyer-Zhao}, {Michalik}, {Monta{\~n}a}, {Montgomerie}, {Mora-Klein}, {Muders}, {Nadolski}, {Navarro}, {Neilsen}, {Nguyen}, {Nishioka}, {Norton}, {Nowak}, {Nystrom}, {Ogawa}, {Oshiro}, {Oyama}, {Parsons}, {Paine}, {Pe{\~n}alver}, {Phillips}, {Poirier}, {Pradel}, {Primiani}, {Raffin}, {Rahlin}, {Reiland}, {Risacher}, {Ruiz}, {S{\'a}ez-Mada{\'\i}n}, {Sassella}, {Schellart}, {Shaw}, {Silva}, {Shiokawa}, {Smith}, {Snow}, {Souccar}, {Sousa}, {Sridharan}, {Srinivasan}, {Stahm}, {Stark}, {Story}, {Timmer}, {Vertatschitsch}, {Walther}, {Wei}, {Whitehorn}, {Whitney}, {Woody}, {Wouterloot}, {Wright}, {Yamaguchi}, {Yu}, {Zeballos}, {Zhang}, \& {Ziurys}}]{EHT2019_M87_paper1}
{Event Horizon Telescope Collaboration}, {Akiyama}, K., {Alberdi}, A., {et~al.} 2019{\natexlab{a}}, \apjl, 875, L1

\bibitem[{{Event Horizon Telescope Collaboration} {et~al.}(2019{\natexlab{b}}){Event Horizon Telescope Collaboration}, {Akiyama}, {Alberdi}, {Alef}, {Asada}, {Azulay}, {Baczko}, {Ball}, {Balokovi{\'c}}, {Barrett}, {Bintley}, {Blackburn}, {Boland}, {Bouman}, {Bower}, {Bremer}, {Brinkerink}, {Brissenden}, {Britzen}, {Broderick}, {Broguiere}, {Bronzwaer}, {Byun}, {Carlstrom}, {Chael}, {Chan}, {Chatterjee}, {Chatterjee}, {Chen}, {Chen}, {Cho}, {Christian}, {Conway}, {Cordes}, {Crew}, {Cui}, {Davelaar}, {De Laurentis}, {Deane}, {Dempsey}, {Desvignes}, {Dexter}, {Doeleman}, {Eatough}, {Falcke}, {Fish}, {Fomalont}, {Fraga-Encinas}, {Friberg}, {Fromm}, {G{\'o}mez}, {Galison}, {Gammie}, {Garc{\'\i}a}, {Gentaz}, {Georgiev}, {Goddi}, {Gold}, {Gu}, {Gurwell}, {Hada}, {Hecht}, {Hesper}, {Ho}, {Ho}, {Honma}, {Huang}, {Huang}, {Hughes}, {Ikeda}, {Inoue}, {Issaoun}, {James}, {Jannuzi}, {Janssen}, {Jeter}, {Jiang}, {Johnson}, {Jorstad}, {Jung}, {Karami}, {Karuppusamy}, {Kawashima}, {Keating}, {Kettenis}, {Kim}, {Kim}, {Kim},
  {Kino}, {Koay}, {Koch}, {Koyama}, {Kramer}, {Kramer}, {Krichbaum}, {Kuo}, {Lauer}, {Lee}, {Li}, {Li}, {Lindqvist}, {Liu}, {Liuzzo}, {Lo}, {Lobanov}, {Loinard}, {Lonsdale}, {Lu}, {MacDonald}, {Mao}, {Markoff}, {Marrone}, {Marscher}, {Mart{\'\i}-Vidal}, {Matsushita}, {Matthews}, {Medeiros}, {Menten}, {Mizuno}, {Mizuno}, {Moran}, {Moriyama}, {Moscibrodzka}, {M{\"u}ller}, {Nagai}, {Nagar}, {Nakamura}, {Narayan}, {Narayanan}, {Natarajan}, {Neri}, {Ni}, {Noutsos}, {Okino}, {Olivares}, {Oyama}, {{\"O}zel}, {Palumbo}, {Patel}, {Pen}, {Pesce}, {Pi{\'e}tu}, {Plambeck}, {PopStefanija}, {Porth}, {Prather}, {Preciado-L{\'o}pez}, {Psaltis}, {Pu}, {Ramakrishnan}, {Rao}, {Rawlings}, {Raymond}, {Rezzolla}, {Ripperda}, {Roelofs}, {Rogers}, {Ros}, {Rose}, {Roshanineshat}, {Rottmann}, {Roy}, {Ruszczyk}, {Ryan}, {Rygl}, {S{\'a}nchez}, {S{\'a}nchez-Arguelles}, {Sasada}, {Savolainen}, {Schloerb}, {Schuster}, {Shao}, {Shen}, {Small}, {Sohn}, {SooHoo}, {Tazaki}, {Tiede}, {Tilanus}, {Titus}, {Toma}, {Torne}, {Trent}, {Trippe},
  {Tsuda}, {van Bemmel}, {van Langevelde}, {van Rossum}, {Wagner}, {Wardle}, {Weintroub}, {Wex}, {Wharton}, {Wielgus}, {Wong}, {Wu}, {Young}, {Young}, {Younsi}, {Yuan}, {Yuan}, {Zensus}, {Zhao}, {Zhao}, {Zhu}, {Farah}, {Meyer-Zhao}, {Michalik}, {Nadolski}, {Nishioka}, {Pradel}, {Primiani}, {Souccar}, {Vertatschitsch}, \& {Yamaguchi}}]{eht2019_paper6_mass}
{Event Horizon Telescope Collaboration}, {Akiyama}, K., {Alberdi}, A., {et~al.} 2019{\natexlab{b}}, \apjl, 875, L6

\bibitem[{{Event Horizon Telescope Collaboration} {et~al.}(2019{\natexlab{c}}){Event Horizon Telescope Collaboration}, {Akiyama}, {Alberdi}, {Alef}, {Asada}, {Azulay}, {Baczko}, {Ball}, {Balokovi{\'c}}, {Barrett}, {Bintley}, {Blackburn}, {Boland}, {Bouman}, {Bower}, {Bremer}, {Brinkerink}, {Brissenden}, {Britzen}, {Broderick}, {Broguiere}, {Bronzwaer}, {Byun}, {Carlstrom}, {Chael}, {Chan}, {Chatterjee}, {Chatterjee}, {Chen}, {Chen}, {Cho}, {Christian}, {Conway}, {Cordes}, {Crew}, {Cui}, {Davelaar}, {De Laurentis}, {Deane}, {Dempsey}, {Desvignes}, {Dexter}, {Doeleman}, {Eatough}, {Falcke}, {Fish}, {Fomalont}, {Fraga-Encinas}, {Freeman}, {Friberg}, {Fromm}, {G{\'o}mez}, {Galison}, {Gammie}, {Garc{\'\i}a}, {Gentaz}, {Georgiev}, {Goddi}, {Gold}, {Gu}, {Gurwell}, {Hada}, {Hecht}, {Hesper}, {Ho}, {Ho}, {Honma}, {Huang}, {Huang}, {Hughes}, {Ikeda}, {Inoue}, {Issaoun}, {James}, {Jannuzi}, {Janssen}, {Jeter}, {Jiang}, {Johnson}, {Jorstad}, {Jung}, {Karami}, {Karuppusamy}, {Kawashima}, {Keating}, {Kettenis}, {Kim},
  {Kim}, {Kim}, {Kino}, {Koay}, {Koch}, {Koyama}, {Kramer}, {Kramer}, {Krichbaum}, {Kuo}, {Lauer}, {Lee}, {Li}, {Li}, {Lindqvist}, {Liu}, {Liuzzo}, {Lo}, {Lobanov}, {Loinard}, {Lonsdale}, {Lu}, {MacDonald}, {Mao}, {Markoff}, {Marrone}, {Marscher}, {Mart{\'\i}-Vidal}, {Matsushita}, {Matthews}, {Medeiros}, {Menten}, {Mizuno}, {Mizuno}, {Moran}, {Moriyama}, {Moscibrodzka}, {M{\"u}ller}, {Nagai}, {Nagar}, {Nakamura}, {Narayan}, {Narayanan}, {Natarajan}, {Neri}, {Ni}, {Noutsos}, {Okino}, {Olivares}, {Oyama}, {{\"O}zel}, {Palumbo}, {Patel}, {Pen}, {Pesce}, {Pi{\'e}tu}, {Plambeck}, {PopStefanija}, {Porth}, {Prather}, {Preciado-L{\'o}pez}, {Psaltis}, {Pu}, {Ramakrishnan}, {Rao}, {Rawlings}, {Raymond}, {Rezzolla}, {Ripperda}, {Roelofs}, {Rogers}, {Ros}, {Rose}, {Roshanineshat}, {Rottmann}, {Roy}, {Ruszczyk}, {Ryan}, {Rygl}, {S{\'a}nchez}, {S{\'a}nchez-Arguelles}, {Sasada}, {Savolainen}, {Schloerb}, {Schuster}, {Shao}, {Shen}, {Small}, {Sohn}, {SooHoo}, {Tazaki}, {Tiede}, {Tilanus}, {Titus}, {Toma}, {Torne}, {Trent},
  {Trippe}, {Tsuda}, {van Bemmel}, {van Langevelde}, {van Rossum}, {Wagner}, {Wardle}, {Weintroub}, {Wex}, {Wharton}, {Wielgus}, {Wong}, {Wu}, {Young}, {Young}, {Younsi}, {Yuan}, {Yuan}, {Zensus}, {Zhao}, {Zhao}, {Zhu}, {Farah}, {Meyer-Zhao}, {Michalik}, {Nadolski}, {Nishioka}, {Pradel}, {Primiani}, {Souccar}, {Vertatschitsch}, \& {Yamaguchi}}]{eht2019_L4_imaging}
{Event Horizon Telescope Collaboration}, {Akiyama}, K., {Alberdi}, A., {et~al.} 2019{\natexlab{c}}, \apjl, 875, L4

\bibitem[{{Event Horizon Telescope Collaboration} {et~al.}(2019{\natexlab{d}}){Event Horizon Telescope Collaboration}, {Akiyama}, {Alberdi}, {Alef}, {Asada}, {Azulay}, {Baczko}, {Ball}, {Balokovi{\'c}}, {Barrett}, {Bintley}, {Blackburn}, {Boland}, {Bouman}, {Bower}, {Bremer}, {Brinkerink}, {Brissenden}, {Britzen}, {Broderick}, {Broguiere}, {Bronzwaer}, {Byun}, {Carlstrom}, {Chael}, {Chan}, {Chatterjee}, {Chatterjee}, {Chen}, {Chen}, {Cho}, {Christian}, {Conway}, {Cordes}, {Crew}, {Cui}, {Davelaar}, {De Laurentis}, {Deane}, {Dempsey}, {Desvignes}, {Dexter}, {Doeleman}, {Eatough}, {Falcke}, {Fish}, {Fomalont}, {Fraga-Encinas}, {Friberg}, {Fromm}, {G{\'o}mez}, {Galison}, {Gammie}, {Garc{\'\i}a}, {Gentaz}, {Georgiev}, {Goddi}, {Gold}, {Gu}, {Gurwell}, {Hada}, {Hecht}, {Hesper}, {Ho}, {Ho}, {Honma}, {Huang}, {Huang}, {Hughes}, {Ikeda}, {Inoue}, {Issaoun}, {James}, {Jannuzi}, {Janssen}, {Jeter}, {Jiang}, {Johnson}, {Jorstad}, {Jung}, {Karami}, {Karuppusamy}, {Kawashima}, {Keating}, {Kettenis}, {Kim}, {Kim}, {Kim},
  {Kino}, {Koay}, {Koch}, {Koyama}, {Kramer}, {Kramer}, {Krichbaum}, {Kuo}, {Lauer}, {Lee}, {Li}, {Li}, {Lindqvist}, {Liu}, {Liuzzo}, {Lo}, {Lobanov}, {Loinard}, {Lonsdale}, {Lu}, {MacDonald}, {Mao}, {Markoff}, {Marrone}, {Marscher}, {Mart{\'\i}-Vidal}, {Matsushita}, {Matthews}, {Medeiros}, {Menten}, {Mizuno}, {Mizuno}, {Moran}, {Moriyama}, {Moscibrodzka}, {Mul{\ensuremath{\ddot{}}}ler}, {Nagai}, {Nagar}, {Nakamura}, {Narayan}, {Narayanan}, {Natarajan}, {Neri}, {Ni}, {Noutsos}, {Okino}, {Olivares}, {Oyama}, {{\"O}zel}, {Palumbo}, {Patel}, {Pen}, {Pesce}, {Pi{\'e}tu}, {Plambeck}, {PopStefanija}, {Porth}, {Prather}, {Preciado-L{\'o}pez}, {Psaltis}, {Pu}, {Ramakrishnan}, {Rao}, {Rawlings}, {Raymond}, {Rezzolla}, {Ripperda}, {Roelofs}, {Rogers}, {Ros}, {Rose}, {Roshanineshat}, {Rottmann}, {Roy}, {Ruszczyk}, {Ryan}, {Rygl}, {S{\'a}nchez}, {S{\'a}nchez-Arguelles}, {Sasada}, {Savolainen}, {Schloerb}, {Schuster}, {Shao}, {Shen}, {Small}, {Sohn}, {SooHoo}, {Tazaki}, {Tiede}, {Tilanus}, {Titus}, {Toma}, {Torne},
  {Trent}, {Trippe}, {Tsuda}, {van Bemmel}, {van Langevelde}, {van Rossum}, {Wagner}, {Wardle}, {Weintroub}, {Wex}, {Wharton}, {Wielgus}, {Wong}, {Wu}, {Young}, {Young}, {Younsi}, {Yuan}, {Yuan}, {Zensus}, {Zhao}, {Zhao}, {Zhu}, {Anczarski}, {Baganoff}, {Eckart}, {Farah}, {Haggard}, {Meyer-Zhao}, {Michalik}, {Nadolski}, {Neilsen}, {Nishioka}, {Nowak}, {Pradel}, {Primiani}, {Souccar}, {Vertatschitsch}, {Yamaguchi}, \& {Zhang}}]{EHT2019_P5}
{Event Horizon Telescope Collaboration}, {Akiyama}, K., {Alberdi}, A., {et~al.} 2019{\natexlab{d}}, \apjl, 875, L5

\bibitem[{{Event Horizon Telescope Collaboration} {et~al.}(2021{\natexlab{a}}){Event Horizon Telescope Collaboration}, {Akiyama}, {Algaba}, {Alberdi}, {Alef}, {Anantua}, {Asada}, {Azulay}, {Baczko}, {Ball}, {Balokovi{\'c}}, {Barrett}, {Benson}, {Bintley}, {Blackburn}, {Blundell}, {Boland}, {Bouman}, {Bower}, {Boyce}, {Bremer}, {Brinkerink}, {Brissenden}, {Britzen}, {Broderick}, {Broguiere}, {Bronzwaer}, {Byun}, {Carlstrom}, {Chael}, {Chan}, {Chatterjee}, {Chatterjee}, {Chen}, {Chen}, {Chesler}, {Cho}, {Christian}, {Conway}, {Cordes}, {Crawford}, {Crew}, {Cruz-Osorio}, {Cui}, {Davelaar}, {De Laurentis}, {Deane}, {Dempsey}, {Desvignes}, {Dexter}, {Doeleman}, {Eatough}, {Falcke}, {Farah}, {Fish}, {Fomalont}, {Ford}, {Fraga-Encinas}, {Freeman}, {Friberg}, {Fromm}, {Fuentes}, {Galison}, {Gammie}, {Garc{\'\i}a}, {Gentaz}, {Georgiev}, {Goddi}, {Gold}, {G{\'o}mez}, {G{\'o}mez-Ruiz}, {Gu}, {Gurwell}, {Hada}, {Haggard}, {Hecht}, {Hesper}, {Ho}, {Ho}, {Honma}, {Huang}, {Huang}, {Hughes}, {Ikeda}, {Inoue}, {Issaoun},
  {James}, {Jannuzi}, {Janssen}, {Jeter}, {Jiang}, {Jimenez-Rosales}, {Johnson}, {Jorstad}, {Jung}, {Karami}, {Karuppusamy}, {Kawashima}, {Keating}, {Kettenis}, {Kim}, {Kim}, {Kim}, {Kim}, {Kino}, {Koay}, {Kofuji}, {Koch}, {Koyama}, {Kramer}, {Kramer}, {Krichbaum}, {Kuo}, {Lauer}, {Lee}, {Levis}, {Li}, {Li}, {Lindqvist}, {Lico}, {Lindahl}, {Liu}, {Liu}, {Liuzzo}, {Lo}, {Lobanov}, {Loinard}, {Lonsdale}, {Lu}, {MacDonald}, {Mao}, {Marchili}, {Markoff}, {Marrone}, {Marscher}, {Mart{\'\i}-Vidal}, {Matsushita}, {Matthews}, {Medeiros}, {Menten}, {Mizuno}, {Mizuno}, {Moran}, {Moriyama}, {Moscibrodzka}, {M{\"u}ller}, {Musoke}, {Mej{\'\i}as}, {Michalik}, {Nadolski}, {Nagai}, {Nagar}, {Nakamura}, {Narayan}, {Narayanan}, {Natarajan}, {Nathanail}, {Neilsen}, {Neri}, {Ni}, {Noutsos}, {Nowak}, {Okino}, {Olivares}, {Ortiz-Le{\'o}n}, {Oyama}, {{\"O}zel}, {Palumbo}, {Park}, {Patel}, {Pen}, {Pesce}, {Pi{\'e}tu}, {Plambeck}, {PopStefanija}, {Porth}, {P{\"o}tzl}, {Prather}, {Preciado-L{\'o}pez}, {Psaltis}, {Pu}, {Ramakrishnan},
  {Rao}, {Rawlings}, {Raymond}, {Rezzolla}, {Ricarte}, {Ripperda}, {Roelofs}, {Rogers}, {Ros}, {Rose}, {Roshanineshat}, {Rottmann}, {Roy}, {Ruszczyk}, {Rygl}, {S{\'a}nchez}, {S{\'a}nchez-Arguelles}, {Sasada}, {Savolainen}, {Schloerb}, {Schuster}, {Shao}, {Shen}, {Small}, {Sohn}, {SooHoo}, {Sun}, {Tazaki}, {Tetarenko}, {Tiede}, {Tilanus}, {Titus}, {Toma}, {Torne}, {Trent}, {Traianou}, {Trippe}, {van Bemmel}, {van Langevelde}, {van Rossum}, {Wagner}, {Ward-Thompson}, {Wardle}, {Weintroub}, {Wex}, {Wharton}, {Wielgus}, {Wong}, {Wu}, {Yoon}, {Young}, {Young}, {Younsi}, {Yuan}, {Yuan}, {Zensus}, {Zhao}, \& {Zhao}}]{eht2021a}
{Event Horizon Telescope Collaboration}, {Akiyama}, K., {Algaba}, J.~C., {et~al.} 2021{\natexlab{a}}, \apjl, 910, L12

\bibitem[{{Event Horizon Telescope Collaboration} {et~al.}(2021{\natexlab{b}}){Event Horizon Telescope Collaboration}, {Akiyama}, {Algaba}, {Alberdi}, {Alef}, {Anantua}, {Asada}, {Azulay}, {Baczko}, {Ball}, {Balokovi{\'c}}, {Barrett}, {Benson}, {Bintley}, {Blackburn}, {Blundell}, {Boland}, {Bouman}, {Bower}, {Boyce}, {Bremer}, {Brinkerink}, {Brissenden}, {Britzen}, {Broderick}, {Broguiere}, {Bronzwaer}, {Byun}, {Carlstrom}, {Chael}, {Chan}, {Chatterjee}, {Chatterjee}, {Chen}, {Chen}, {Chesler}, {Cho}, {Christian}, {Conway}, {Cordes}, {Crawford}, {Crew}, {Cruz-Osorio}, {Cui}, {Davelaar}, {De Laurentis}, {Deane}, {Dempsey}, {Desvignes}, {Dexter}, {Doeleman}, {Eatough}, {Falcke}, {Farah}, {Fish}, {Fomalont}, {Ford}, {Fraga-Encinas}, {Friberg}, {Fromm}, {Fuentes}, {Galison}, {Gammie}, {Garc{\'\i}a}, {Gelles}, {Gentaz}, {Georgiev}, {Goddi}, {Gold}, {G{\'o}mez}, {G{\'o}mez-Ruiz}, {Gu}, {Gurwell}, {Hada}, {Haggard}, {Hecht}, {Hesper}, {Himwich}, {Ho}, {Ho}, {Honma}, {Huang}, {Huang}, {Hughes}, {Ikeda}, {Inoue},
  {Issaoun}, {James}, {Jannuzi}, {Janssen}, {Jeter}, {Jiang}, {Jimenez-Rosales}, {Johnson}, {Jorstad}, {Jung}, {Karami}, {Karuppusamy}, {Kawashima}, {Keating}, {Kettenis}, {Kim}, {Kim}, {Kim}, {Kim}, {Kino}, {Koay}, {Kofuji}, {Koch}, {Koyama}, {Kramer}, {Kramer}, {Krichbaum}, {Kuo}, {Lauer}, {Lee}, {Levis}, {Li}, {Li}, {Lindqvist}, {Lico}, {Lindahl}, {Liu}, {Liu}, {Liuzzo}, {Lo}, {Lobanov}, {Loinard}, {Lonsdale}, {Lu}, {MacDonald}, {Mao}, {Marchili}, {Markoff}, {Marrone}, {Marscher}, {Mart{\'\i}-Vidal}, {Matsushita}, {Matthews}, {Medeiros}, {Menten}, {Mizuno}, {Mizuno}, {Moran}, {Moriyama}, {Moscibrodzka}, {M{\"u}ller}, {Musoke}, {Mus Mej{\'\i}as}, {Michalik}, {Nadolski}, {Nagai}, {Nagar}, {Nakamura}, {Narayan}, {Narayanan}, {Natarajan}, {Nathanail}, {Neilsen}, {Neri}, {Ni}, {Noutsos}, {Nowak}, {Okino}, {Olivares}, {Ortiz-Le{\'o}n}, {Oyama}, {{\"O}zel}, {Palumbo}, {Park}, {Patel}, {Pen}, {Pesce}, {Pi{\'e}tu}, {Plambeck}, {PopStefanija}, {Porth}, {P{\"o}tzl}, {Prather}, {Preciado-L{\'o}pez}, {Psaltis}, {Pu},
  {Ramakrishnan}, {Rao}, {Rawlings}, {Raymond}, {Rezzolla}, {Ricarte}, {Ripperda}, {Roelofs}, {Rogers}, {Ros}, {Rose}, {Roshanineshat}, {Rottmann}, {Roy}, {Ruszczyk}, {Rygl}, {S{\'a}nchez}, {S{\'a}nchez-Arguelles}, {Sasada}, {Savolainen}, {Schloerb}, {Schuster}, {Shao}, {Shen}, {Small}, {Sohn}, {SooHoo}, {Sun}, {Tazaki}, {Tetarenko}, {Tiede}, {Tilanus}, {Titus}, {Toma}, {Torne}, {Trent}, {Traianou}, {Trippe}, {van Bemmel}, {van Langevelde}, {van Rossum}, {Wagner}, {Ward-Thompson}, {Wardle}, {Weintroub}, {Wex}, {Wharton}, {Wielgus}, {Wong}, {Wu}, {Yoon}, {Young}, {Young}, {Younsi}, {Yuan}, {Yuan}, {Zensus}, {Zhao}, \& {Zhao}}]{eht2021b}
{Event Horizon Telescope Collaboration}, {Akiyama}, K., {Algaba}, J.~C., {et~al.} 2021{\natexlab{b}}, \apjl, 910, L13

\bibitem[{{Frank} {et~al.}(2021){Frank}, {Leike}, \& {En{\ss}lin}}]{Frank2021}
{Frank}, P., {Leike}, R., \& {En{\ss}lin}, T.~A. 2021, Entropy, 23, 853

\bibitem[{{Georgiev} {et~al.}(2022){Georgiev}, {Pesce}, {Broderick}, {Wong}, {Dhruv}, {Wielgus}, {Gammie}, {Chan}, {Chatterjee}, {Emami}, {Mizuno}, {Gold}, {Fromm}, {Ricarte}, {Yoon}, {Joshi}, {Prather}, {Cruz-Osorio}, {Johnson}, {Porth}, {Olivares}, {Younsi}, {Rezzolla}, {Vos}, {Qiu}, {Nathanail}, {Narayan}, {Chael}, {Anantua}, {Moscibrodzka}, {Akiyama}, {Alberdi}, {Alef}, {Algaba}, {Asada}, {Azulay}, {Bach}, {Baczko}, {Ball}, {Balokovi{\'c}}, {Barrett}, {Baub{\"o}ck}, {Benson}, {Bintley}, {Blackburn}, {Blundell}, {Bouman}, {Bower}, {Boyce}, {Bremer}, {Brinkerink}, {Brissenden}, {Britzen}, {Broguiere}, {Bronzwaer}, {Bustamante}, {Byun}, {Carlstrom}, {Ceccobello}, {Chatterjee}, {Chen}, {Chen}, {Cheng}, {Cho}, {Christian}, {Conroy}, {Conway}, {Cordes}, {Crawford}, {Crew}, {Cui}, {Davelaar}, {De Laurentis}, {Deane}, {Dempsey}, {Desvignes}, {Dexter}, {Doeleman}, {Dougal}, {Dzib}, {Eatough}, {Falcke}, {Farah}, {Fish}, {Fomalont}, {Ford}, {Fraga-Encinas}, {Freeman}, {Friberg}, {Fuentes}, {Galison}, {Garc{\'\i}a},
  {Gentaz}, {Goddi}, {G{\'o}mez-Ruiz}, {G{\'o}mez}, {Gu}, {Gurwell}, {Hada}, {Haggard}, {Haworth}, {Hecht}, {Hesper}, {Heumann}, {Ho}, {Ho}, {Honma}, {Huang}, {Huang}, {Hughes}, {Ikeda}, {Impellizzeri}, {Inoue}, {Issaoun}, {James}, {Jannuzi}, {Janssen}, {Jeter}, {Jiang}, {Jim{\'e}nez-Rosales}, {Jorstad}, {Jung}, {Karami}, {Karuppusamy}, {Kawashima}, {Keating}, {Kettenis}, {Kim}, {Kim}, {Kim}, {Kim}, {Kino}, {Koay}, {Kocherlakota}, {Kofuji}, {Koch}, {Koyama}, {Kramer}, {Kramer}, {Krichbaum}, {Kuo}, {La Bella}, {Lauer}, {Lee}, {Lee}, {Lehner}, {Leung}, {Levis}, {Li}, {Lico}, {Lindahl}, {Lindqvist}, {Lisakov}, {Liu}, {Liu}, {Liuzzo}, {Lo}, {Lobanov}, {Loinard}, {Lonsdale}, {Lu}, {Mao}, {Marchili}, {Markoff}, {Marrone}, {Marscher}, {Mart{\'\i}-Vidal}, {Matsushita}, {Matthews}, {Menten}, {Michalik}, {Mizuno}, {Moran}, {Moriyama}, {M{\"u}ller}, {Mus}, {Musoke}, {Myserlis}, {Nadolski}, {Nagai}, {Nagar}, {Nakamura}, {Narayanan}, {Natarajan}, {Navarro Fuentes}, {Neilsen}, {Neri}, {Ni}, {Noutsos}, {Nowak}, {Oh},
  {Okino}, {Ortiz-Le{\'o}n}, {Oyama}, {Palumbo}, {Paraschos}, {Park}, {Parsons}, {Patel}, {Pen}, {Pi{\'e}tu}, {Plambeck}, {PopStefanija}, {P{\"o}tzl}, {Preciado-L{\'o}pez}, {Pu}, {Ramakrishnan}, {Rao}, {Rawlings}, {Raymond}, {Ripperda}, {Roelofs}, {Rogers}, {Ros}, {Romero-Ca{\~n}izales}, {Roshanineshat}, {Rottmann}, {Roy}, {Ruiz}, {Ruszczyk}, {Rygl}, {S{\'a}nchez}, {S{\'a}nchez-Arg{\"u}elles}, {S{\'a}nchez-Portal}, {Sasada}, {Satapathy}, {Savolainen}, {Schloerb}, {Schonfeld}, {Schuster}, {Shao}, {Shen}, {Small}, {Sohn}, {SooHoo}, {Souccar}, {Sun}, {Tazaki}, {Tetarenko}, {Tiede}, {Tilanus}, {Titus}, {Torne}, {Traianou}, {Trent}, {Trippe}, {Turk}, {van Bemmel}, {van Langevelde}, {van Rossum}, {Wagner}, {Ward-Thompson}, {Wardle}, {Weintroub}, {Wex}, {Wharton}, {Wiik}, {Witzel}, {Wondrak}, {Wu}, {Yamaguchi}, {Young}, {Young}, {Yuan}, {Yuan}, {Zensus}, {Zhang}, {Zhao}, \& {Zhao}}]{Georgiev2022}
{Georgiev}, B., {Pesce}, D.~W., {Broderick}, A.~E., {et~al.} 2022, \apjl, 930, L20

\bibitem[{{Hada}(2017)}]{2017Galax...5....2H}
{Hada}, K. 2017, Galaxies, 5, 2

\bibitem[{{H{\"o}gbom}(1974)}]{Hogbom1974}
{H{\"o}gbom}, J.~A. 1974, \aaps, 15, 417

\bibitem[{{Issaoun} {et~al.}(2019){Issaoun}, {Johnson}, {Blackburn}, {Brinkerink}, {Mo{\'s}cibrodzka}, {Chael}, {Goddi}, {Mart{\'\i}-Vidal}, {Wagner}, {Doeleman}, {Falcke}, {Krichbaum}, {Akiyama}, {Bach}, {Bouman}, {Bower}, {Broderick}, {Cho}, {Crew}, {Dexter}, {Fish}, {Gold}, {G{\'o}mez}, {Hada}, {Hern{\'a}ndez-G{\'o}mez}, {Jan{\ss}en}, {Kino}, {Kramer}, {Loinard}, {Lu}, {Markoff}, {Marrone}, {Matthews}, {Moran}, {M{\"u}ller}, {Roelofs}, {Ros}, {Rottmann}, {Sanchez}, {Tilanus}, {de Vicente}, {Wielgus}, {Zensus}, \& {Zhao}}]{Issaoun2019}
{Issaoun}, S., {Johnson}, M.~D., {Blackburn}, L., {et~al.} 2019, \apj, 871, 30

\bibitem[{{Junklewitz} {et~al.}(2016){Junklewitz}, {Bell}, {Selig}, \& {En{\ss}lin}}]{Junklewitz2016}
{Junklewitz}, H., {Bell}, M.~R., {Selig}, M., \& {En{\ss}lin}, T.~A. 2016, \aap, 586, A76

\bibitem[{{Kim} {et~al.}(2024){Kim}, {Nikonov}, {Roth}, {En{\ss}lin}, {Janssen}, {Arras}, {M{\"u}ller}, \& {Lobanov}}]{JSKim2024}
{Kim}, J.-S., {Nikonov}, A.~S., {Roth}, J., {et~al.} 2024, \aap, 690, A129

\bibitem[{{Kim} {et~al.}(2018){Kim}, {Krichbaum}, {Lu}, {Ros}, {Bach}, {Bremer}, {de Vicente}, {Lindqvist}, \& {Zensus}}]{JYKim2018}
{Kim}, J.~Y., {Krichbaum}, T.~P., {Lu}, R.~S., {et~al.} 2018, \aap, 616, A188

\bibitem[{{Kim} {et~al.}(2023){Kim}, {Savolainen}, {Voitsik}, {Kravchenko}, {Lisakov}, {Kovalev}, {M{\"u}ller}, {Lobanov}, {Sokolovsky}, {Bruni}, {Edwards}, {Reynolds}, {Bach}, {Gurvits}, {Krichbaum}, {Hada}, {Giroletti}, {Orienti}, {Anderson}, {Lee}, {Sohn}, \& {Zensus}}]{JYKim2023}
{Kim}, J.-Y., {Savolainen}, T., {Voitsik}, P., {et~al.} 2023, \apj, 952, 34

\bibitem[{{Knollm{\"u}ller} \& {En{\ss}lin}(2019)}]{Knollmueller2019}
{Knollm{\"u}ller}, J. \& {En{\ss}lin}, T.~A. 2019, arXiv e-prints, arXiv:1901.11033

\bibitem[{{Lannes} {et~al.}(1997){Lannes}, {Anterrieu}, \& {Marechal}}]{Lannes1997}
{Lannes}, A., {Anterrieu}, E., \& {Marechal}, P. 1997, \aaps, 123, 183

\bibitem[{{Lu} {et~al.}(2023){Lu}, {Asada}, {Krichbaum}, {Park}, {Tazaki}, {Pu}, {Nakamura}, {Lobanov}, {Hada}, {Akiyama}, {Kim}, {Marti-Vidal}, {G{\'o}mez}, {Kawashima}, {Yuan}, {Ros}, {Alef}, {Britzen}, {Bremer}, {Broderick}, {Doi}, {Giovannini}, {Giroletti}, {Ho}, {Honma}, {Hughes}, {Inoue}, {Jiang}, {Kino}, {Koyama}, {Lindqvist}, {Liu}, {Marscher}, {Matsushita}, {Nagai}, {Rottmann}, {Savolainen}, {Schuster}, {Shen}, {de Vicente}, {Walker}, {Yang}, {Zensus}, {Algaba}, {Allardi}, {Bach}, {Berthold}, {Bintley}, {Byun}, {Casadio}, {Chang}, {Chang}, {Chang}, {Chen}, {Chen}, {Chilson}, {Chuter}, {Conway}, {Crew}, {Dempsey}, {Dornbusch}, {Faber}, {Friberg}, {Garc{\'\i}a}, {Garrido}, {Han}, {Han}, {Hasegawa}, {Herrero-Illana}, {Huang}, {Huang}, {Impellizzeri}, {Jiang}, {Jinchi}, {Jung}, {Kallunki}, {Kirves}, {Kimura}, {Koay}, {Koch}, {Kramer}, {Kraus}, {Kubo}, {Kuo}, {Li}, {Lin}, {Liu}, {Liu}, {Lo}, {Lu}, {MacDonald}, {Martin-Cocher}, {Messias}, {Meyer-Zhao}, {Minter}, {Nair}, {Nishioka}, {Norton}, {Nystrom},
  {Ogawa}, {Oshiro}, {Patel}, {Pen}, {Pidopryhora}, {Pradel}, {Raffin}, {Rao}, {Ruiz}, {Sanchez}, {Shaw}, {Snow}, {Sridharan}, {Srinivasan}, {Tercero}, {Torne}, {Traianou}, {Wagner}, {Walther}, {Wei}, {Yang}, \& {Yu}}]{Lu2023}
{Lu}, R.-S., {Asada}, K., {Krichbaum}, T.~P., {et~al.} 2023, \nat, 616, 686

\bibitem[{{M{\"u}ller} \& {Lobanov}(2022)}]{Mueller2022}
{M{\"u}ller}, H. \& {Lobanov}, A.~P. 2022, \aap, 666, A137

\bibitem[{{M{\"u}ller} \& {Lobanov}(2023{\natexlab{a}})}]{Mueller2023a}
{M{\"u}ller}, H. \& {Lobanov}, A.~P. 2023{\natexlab{a}}, \aap, 673, A151

\bibitem[{{M{\"u}ller} \& {Lobanov}(2023{\natexlab{b}})}]{Mueller2023b}
{M{\"u}ller}, H. \& {Lobanov}, A.~P. 2023{\natexlab{b}}, \aap, 672, A26

\bibitem[{{M{\"u}ller} {et~al.}(2024){M{\"u}ller}, {Massa}, {Mus}, {Kim}, \& {Perracchione}}]{Mueller2024_STIX_VLBI}
{M{\"u}ller}, H., {Massa}, P., {Mus}, A., {Kim}, J.-S., \& {Perracchione}, E. 2024, \aap, 684, A47

\bibitem[{{M{\"u}ller} {et~al.}(2023){M{\"u}ller}, {Mus}, \& {Lobanov}}]{Mueller2023c}
{M{\"u}ller}, H., {Mus}, A., \& {Lobanov}, A. 2023, \aap, 675, A60

\bibitem[{{Mus} {et~al.}(2024{\natexlab{a}}){Mus}, {M{\"u}ller}, \& {Lobanov}}]{Mus2024b}
{Mus}, A., {M{\"u}ller}, H., \& {Lobanov}, A. 2024{\natexlab{a}}, \aap, 688, A100

\bibitem[{{Mus} {et~al.}(2024{\natexlab{b}}){Mus}, {M{\"u}ller}, {Mart{\'\i}-Vidal}, \& {Lobanov}}]{Mus2024a}
{Mus}, A., {M{\"u}ller}, H., {Mart{\'\i}-Vidal}, I., \& {Lobanov}, A. 2024{\natexlab{b}}, \aap, 684, A55

\bibitem[{{Nikonov} {et~al.}(2023){Nikonov}, {Kovalev}, {Kravchenko}, {Pashchenko}, \& {Lobanov}}]{Nikonov2023}
{Nikonov}, A.~S., {Kovalev}, Y.~Y., {Kravchenko}, E.~V., {Pashchenko}, I.~N., \& {Lobanov}, A.~P. 2023, \mnras, 526, 5949

\bibitem[{{Okino} {et~al.}(2022){Okino}, {Akiyama}, {Asada}, {G{\'o}mez}, {Hada}, {Honma}, {Krichbaum}, {Kino}, {Nagai}, {Bach}, {Blackburn}, {Bouman}, {Chael}, {Crew}, {Doeleman}, {Fish}, {Goddi}, {Issaoun}, {Johnson}, {Jorstad}, {Koyama}, {Lonsdale}, {Lu}, {Mart{\'\i}-Vidal}, {Matthews}, {Mizuno}, {Moriyama}, {Nakamura}, {Pu}, {Ros}, {Savolainen}, {Tazaki}, {Wagner}, {Wielgus}, \& {Zensus}}]{Okino_2022}
{Okino}, H., {Akiyama}, K., {Asada}, K., {et~al.} 2022, \apj, 940, 65

\bibitem[{{Palumbo} {et~al.}(2024){Palumbo}, {Baub{\"o}ck}, \& {Gammie}}]{Palumbo2024}
{Palumbo}, D. C.~M., {Baub{\"o}ck}, M., \& {Gammie}, C.~F. 2024, \apj, 970, 151

\bibitem[{{Pashchenko} {et~al.}(2023){Pashchenko}, {Kravchenko}, {Nokhrina}, \& {Nikonov}}]{Pashchenko2023}
{Pashchenko}, I.~N., {Kravchenko}, E.~V., {Nokhrina}, E.~E., \& {Nikonov}, A.~S. 2023, \mnras, 523, 1247

\bibitem[{{Roth} {et~al.}(2023){Roth}, {Arras}, {Reinecke}, {Perley}, {Westermann}, \& {En{\ss}lin}}]{Roth2023}
{Roth}, J., {Arras}, P., {Reinecke}, M., {et~al.} 2023, \aap, 678, A177

\bibitem[{{Thyagarajan} {et~al.}(2022){Thyagarajan}, {Nityananda}, \& {Samuel}}]{Thyagarajan2022}
{Thyagarajan}, N., {Nityananda}, R., \& {Samuel}, J. 2022, \prd, 105, 043019

\bibitem[{{Tiede} {et~al.}(2022){Tiede}, {Broderick}, \& {Palumbo}}]{VIDA}
{Tiede}, P., {Broderick}, A.~E., \& {Palumbo}, D. C.~M. 2022, ApJ, 925, 122

\bibitem[{{Twiss} {et~al.}(1960){Twiss}, {Carter}, \& {Little}}]{Twiss1960}
{Twiss}, R.~Q., {Carter}, A.~W.~L., \& {Little}, A.~G. 1960, The Observatory, 80, 153

\bibitem[{{Walker} {et~al.}(2018){Walker}, {Hardee}, {Davies}, {Ly}, \& {Junor}}]{Walker2018}
{Walker}, R.~C., {Hardee}, P.~E., {Davies}, F.~B., {Ly}, C., \& {Junor}, W. 2018, \apj, 855, 128

\bibitem[{{Yang} {et~al.}(2024){Yang}, {Yuan}, {Li}, {Mizuno}, {Guo}, {Lu}, {Ho}, {Lin}, {Zdziarski}, \& {Wang}}]{Yang2024}
{Yang}, H., {Yuan}, F., {Li}, H., {et~al.} 2024, Science Advances, 10, eadn3544

\end{thebibliography}

\appendix

\section{Uncertainty estimation by \texttt{resolve}}

\begin{figure}
    \centering
    \includegraphics[width=9cm]
    {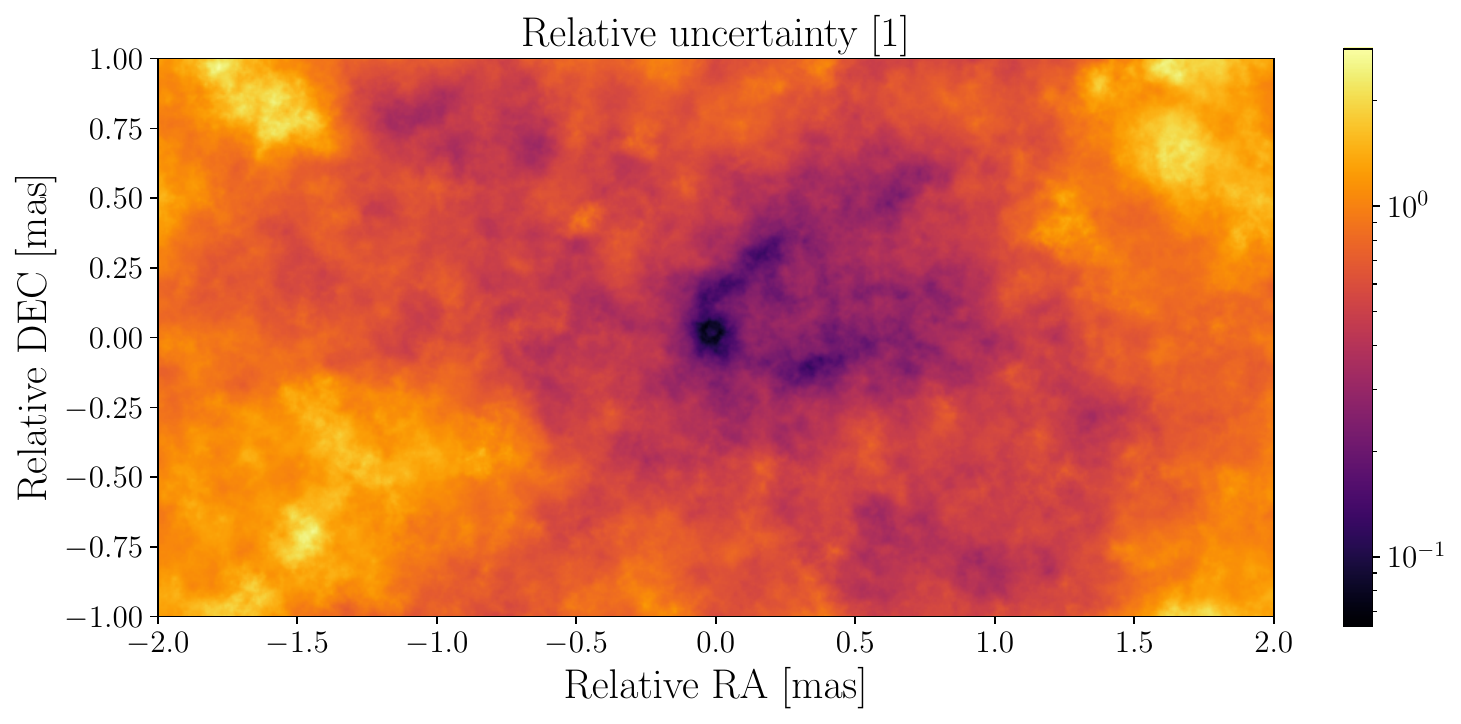}
    \caption{\texttt{resolve} image pixel-wise relative uncertainty, which is the sky  brightness posterior standard deviation normalized by the posterior mean from the \texttt{resolve} reconstruction in the top panel of Fig. \ref{fig:summary}.
    }
    \label{fig:resolve_sky_relative_uncertainty_plot}
\end{figure}

\begin{figure}
    \centering
    \includegraphics[width=9cm]
    {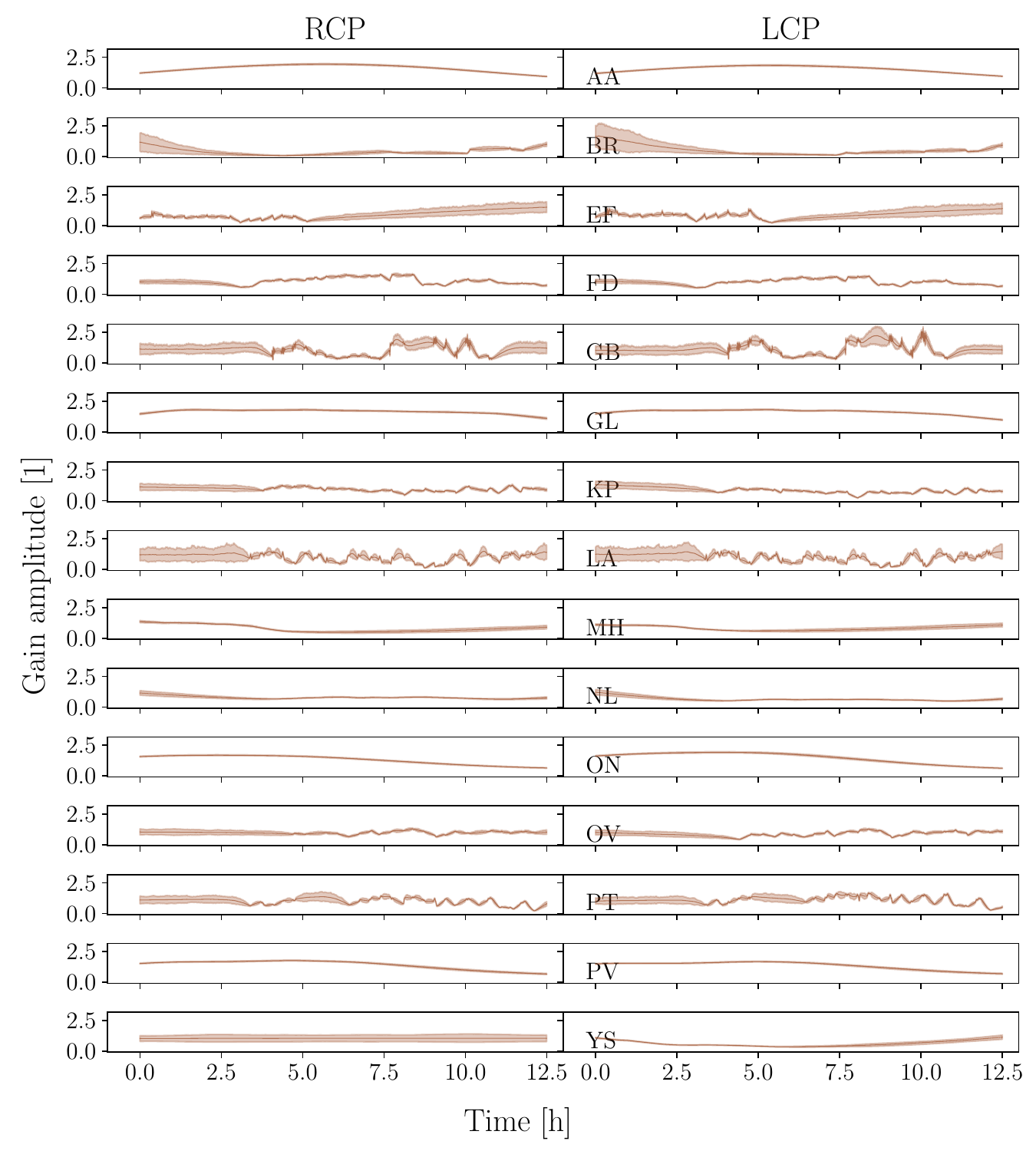}
    \caption{\texttt{resolve} posterior amplitude gains. The gain as a function of time is illustrated as a thin line with a semi-transparent standard deviation. The left and right columns of the figure show gains from the right (RCP) and left (LCP) circular polarizations correspondingly. Each row represents an individual antenna, whose abbreviated name is indicated in the bottom left corner of each LCP plot.
    }
    \label{fig:resolve_amp_gain_plot}
\end{figure}

Figure \ref{fig:resolve_sky_relative_uncertainty_plot} shows the pixel-wise relative uncertainty of the 
\texttt{resolve} image in Fig. \ref{fig:summary}. The relative uncertainty is defined as the sky brightness posterior standard deviation normalized by the posterior mean from 100 posterior sample images. Lower relative uncertainty values of the M87 ring emission and limb-brightened are estimated, that means they are well-constrained by the data. However, the relative uncertainty values in the central spine are higher compared to the limb-brightened jet emission. Figure \ref{fig:resolve_amp_gain_plot} depicts the \texttt{resolve} posterior amplitude gains. ALMA (AA) amplitude gain solutions show smooth behavior with lower posterior standard deviations compared to other arrays due to the high sensitivity of ALMA array. The gain amplitudes with high uncertainties result from the absence of the data (BR: <4h, EF: >6h, GB: <4h, KP: <3h, LA: <3h, PT: <3h, and YS: RCP gains). For instance, the RCP gain amplitudes for YS antenna have uniformly distributed high uncertainties since only single polarization mode (LCP) was observed for YS antenna. 

\section{Contour plots}

\begin{figure*}
    \centering
    \includegraphics[width=18cm, trim={0cm 0cm 0cm 0cm},clip]{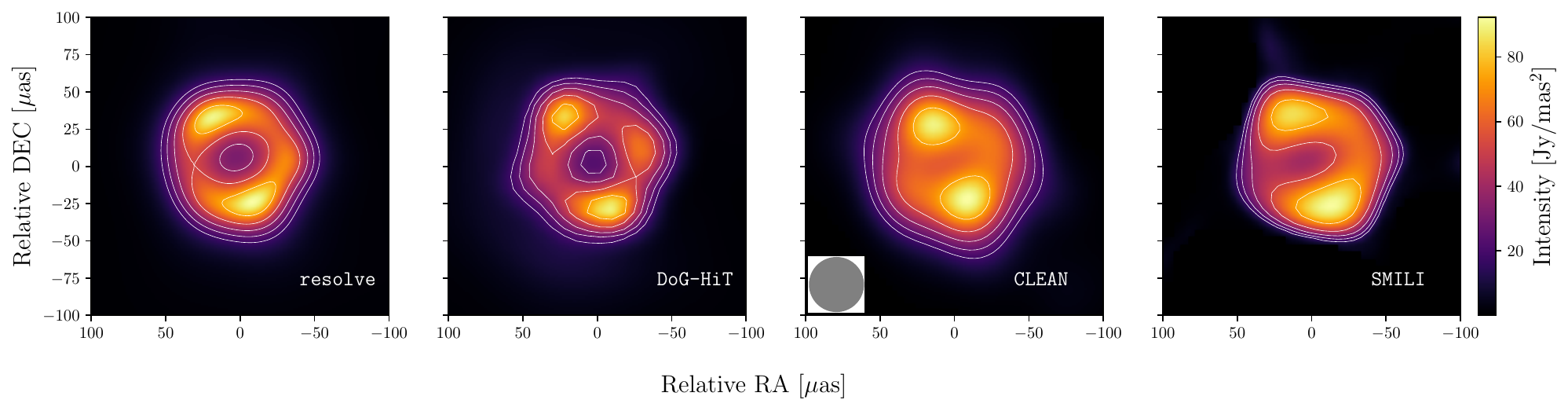}
    \caption{Same as Fig. \ref{fig:ring_compare} , but overlaid with successive contours that increase by a factor of $\sqrt{2}$, starting from 20\% of the peak. 
    } 
    \label{fig:ring_compare_appendix}
\end{figure*}

\begin{figure}
    \centering
    \includegraphics[width=9cm, trim={0cm 0cm 0cm 0cm},clip]{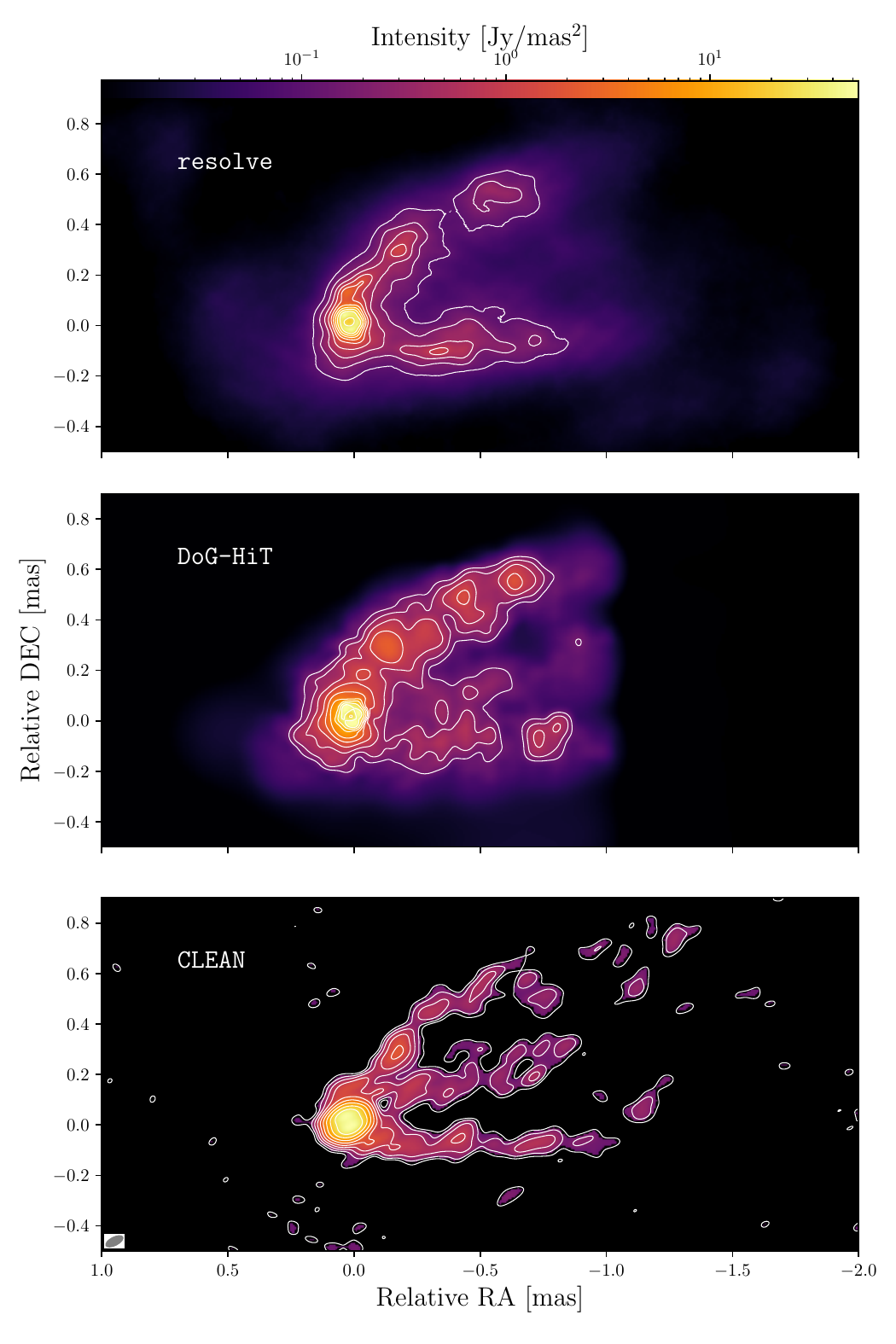}
    \caption{Same as Fig. \ref{fig:summary}, but overlaid with successive contours that increase by a factor of 2, starting from 0.2\% of the peak. The starting level was chosen based on the structure of the \texttt{CLEAN} image.} 
    \label{fig:summary_appendix}
\end{figure}

\end{document}